\documentclass[10pt,letterpaper]{article}

\usepackage[top=0.85in,bottom=0.85in,left=1.25in,right=1.25in,footskip=0.75in]{geometry}

\usepackage[utf8]{inputenc}
\usepackage[T1]{fontenc}

\usepackage{amsmath,amssymb,amsfonts}

\usepackage{microtype} % Handles micro-justification to prevent awkward spacing/hyphenation

\usepackage{changepage}
\usepackage{adjustbox}
\usepackage{wrapfig}

\usepackage{graphicx}
\usepackage[aboveskip=1pt,labelfont=bf,labelsep=period,singlelinecheck=off]{caption}

\usepackage{xcolor}
\definecolor{Gray}{gray}{.25}

\usepackage{nameref}
\usepackage[colorlinks=true, linkcolor=blue, citecolor=blue, urlcolor=blue]{hyperref}

\usepackage{cite}

\usepackage[left]{lineno}

\makeatletter
\renewcommand{\@biblabel}[1]{\quad#1.}
\makeatother

\usepackage{lastpage,fancyhdr}
\begin{document}
\vspace*{0.35in}

% title goes here:
\begin{flushleft}
{\Large
\textbf\newline{MAPCast: A Convection Allowing MPAS Emulator for Ensemble-based Background Error Covariance Estimation Toward Multi-Scale Data Assimilation \footnote{This manuscript has not yet been peer-reviewed and is provided by the contributing Authors as a means to ensure timely dissemination of scholarly and technical work on a noncommercial basis. Copyright and all rights therein are maintained by the Authors or by other copyright owners. It is understood that all persons copying this information will adhere to the terms and constraints invoked by each Author's copyright. This manuscript may not be reposted without explicit permission of the copyright owner.}}
}
\newline
% authors go here:
\\
Yongming Wang\textsuperscript{1,2},
Xuguang Wang\textsuperscript{1,2}
\\
\bigskip
\bf{1} School of Meteorology, University of Oklahoma, Norman, Oklahoma
\\
\bf{2} Consortium for Advanced Data assimilation Research and Education (CADRE)
\\
\bigskip

\end{flushleft}

% now start line numbers
%\linenumbers

\section*{Abstract}
Machine learning (ML) emulators offer a cost-efficient alternative to numerical weather prediction models for generating convection-allowing background ensembles in ensemble-based data assimilation (DA). However, few studies have explored ML-based surrogate background ensembles for estimating background-error covariances (BECs). This study develops a convection-allowing emulator, MAPCast, trained on historical convection-allowing simulations from the Model for Prediction Across Scales (MPAS), and evaluates its ability to estimate BECs, paving the way toward multiscale DA. The evaluation uses 10 retrospective convective cases at 15- and 60-min forecast lead times corresponding to subhourly and hourly DA. MAPCast reproduces MPAS forecasts with good fidelity, including realistic storm coverage, temporal evolution, and similar spatial and spectral characteristics of state variables. Discrepancies are primarily confined to small spatial scales near sharp gradients and convective-scale features and variables. For BEC statistics, MAPCast captures ensemble spread magnitude and spatial distribution for most variables, although larger errors occur for storm-related fields that are vertical velocity and reflectivity. Correlation structures are reproduced most faithfully at mesoscale and above, followed by at convective scales, whereas cross-variable correlations are less accurately represented than univariate correlations, indicating that multivariate coupling remains the principal limitation. MAPCast shows weaker replication of full-scale versus decomposed large and small-scale correlations. BEC estimates derived from 15-min forecasts consistently outperform those from 60-min forecasts, suggesting that shorter lead times better preserve flow-dependent error structures.

% the * after section prevents numbering
\section*{Introduction}

Recent advances in machine learning (ML) emulators have rapidly narrowed the gap with traditional numerical weather prediction (NWP) models, offering high-fidelity forecasts at greatly reduced computational cost. Transformer- and Fourier-based architectures have demonstrated global medium-range forecasting capabilities comparable to state-of-the-art NWP models, as evidenced by FourCastNet \cite{pathakFourCastNetGlobalDatadriven2022}. Subsequent developments, such as PanGu-Weather \cite{biAccurateMediumrangeGlobal2023} and GraphCast \cite{lamLearningSkillfulMediumrange2023}, have further improved forecast accuracy, stability, and large-scale physical consistency through Earth-specific transformer designs and graph neural network (GNN) formulations, respectively. Evaluations have shown competitive performance using multiple large-scale metrics \cite{hakimDynamicalTestsDeep2024}. Growing attention has recently shifted toward regional and convection-allowing applications. For example, Neural-LAM \cite{oskarssonGraphbasedNeuralWeather2023} extends GraphCast to limited-area dynamics. StormCast \cite{pathakKilometerscaleConvectionallowingModel2026} introduces a generative diffusion framework to emulate the 3-km High-Resolution Rapid Refresh (HRRR) analyses \cite{dowellHighResolutionRapidRefresh2022,jamesHighResolutionRapidRefresh2022}. Additionally, WoFsCast \cite{floraWoFSCastMachineLearning2025} adapts GraphCast to the Warn-on-Forecast system (WoFs), and HRRRCast \cite{abdiHRRRCastDataDrivenEmulator2026} explores emulating HRRR analyses using residual network (ResNet) and GNN formulations. Building on earlier ML-based forecasting efforts \cite{scherDataDrivenWeatherClimate2018,weynImprovingDataDrivenGlobal2020,raspDataDrivenMediumRangeWeather2021}, these developments collectively show that ML models, often at orders-of-magnitude lower computational cost, can reproduce many aspects of deterministic or ensemble forecasts over lead times ranging from hours to days. 

The development of ML emulators in previous work has largely emphasized free-forecast performance. However, a fundamental and unresolved question remains: Can ML emulators reliably replace short-term NWP forecasts as background ensemble generators for estimating flow-dependent background-error covariances (BECs) in ensemble-based data assimilation (DA) systems? In conventional ensemble DA, BECs are estimated from physics-based ensembles generated by computationally expensive NWP models. The primary appeal of ML emulators is their computational efficiency, which in principle enables the generation of substantially larger ensembles at negligible additional cost \cite{pathakFourCastNetGlobalDatadriven2022,pathakKilometerscaleConvectionallowingModel2026,lamLearningSkillfulMediumrange2023,biAccurateMediumrangeGlobal2023}. However, this advantage is meaningful only if the surrogate background ensembles maintain realistic ensemble spread and physically consistent correlation structures. High free-forecast skill alone does not guarantee an accurate representation of short-term error growth and flow-dependent correlations, both of which are essential for effective BEC estimation in ensemble DA. Therefore, it is necessary to evaluate the capability of ML surrogate background ensemble in estimating the BECs.

Initial evaluations of ML surrogate background ensembles have primarily focused on global, coarse-resolution ensemble-based DA systems. For example, \cite{liFuXiEn4DVarAssimilationSystem2024} examined the contribution of surrogate BECs generated by the FuXi ML model at a grid spacing of 0.25° within an ensemble four-dimensional variational (En4DVar) framework, while \cite{slivinskiAssimilatingObservedSurface2025} evaluated surrogate background ensembles from various global emulators at resolutions of up to 0.25° in cycling ensemble Kalman filter systems. Although these studies offered initial efforts to understand the effectiveness of the surrogate background ensembles for representing large-scale BECs, challenge increases significantly for convection-allowing applications. Convection-allowing surrogates need explicitly resolve rapidly evolving, nonlinear, and intermittently forced processes, including storm-scale updrafts, cold pools, and localized precipitation extremes. They are required to represent individual storms, the larger-scale supporting environment and their interactions \cite{fabryWhyAreRadar2020,flournoyInfluenceGroundRelativeFlow2021,lombardoProcessesControllingStructure2013,schumacherConvectiveStormEnvironmentsSubtropical2021,sodhiBenefitsSmoothingBackgrounds2022,workoffInfluenceLakeErie2012,wangSimultaneousMultiscaleData2023,wangImprovingCONUSConvectiveScale2024}. Additionally, to qualify as the surrogate background ensemble for convection-allowing model (CAM) DA, they need to account for error growth that is strongly scale dependent and highly anisotropic and provide physically consistent error covariance structures linking larger scales and convective processes. In other words, surrogate background ensembles for CAM DA need much more than just reproducing traditional forecast statistics.  Achieving the aforementioned BEC fidelity for CAM ensemble DA with a purely data-driven emulator is nontrivial and largely unexplored. To date, development and rigorous evaluation of convection-allowing ML emulators explicitly designed for surrogating ensemble BEC in high-resolution DA that resolves multiple scales remain limited.

Substantial challenges remain before the convection-allowing surrogate background ensembles can be effectively integrated into DA systems that update atmospheric states across multiple spatial and temporal scales. For instance, StormCast \cite{pathakKilometerscaleConvectionallowingModel2026} produces forecasts at hourly intervals, making it unsuitable for subhourly cycling DA systems (e.g., WoFs; \cite{stensrudConvectiveScaleWarnonForecastSystem2009,stensrudProgressChallengesWarnonForecast2013,lawsonAdvancingConvectionAllowingNWP2018,skinnerObjectBasedVerificationPrototype2018}), which are designed to capture the rapidly evolving storms.  Although WoFsCast \cite{floraWoFSCastMachineLearning2025} provides forecasts at subhourly resolution, its limited-area domain mostly resolves storm-scale processes and does not adequately represent the mesoscale environment that governs storm initiation and evolution. Another important limitation is that existing convection-allowing emulators often omit the forecast of critical variables essential for convective scale DA, most notably 3-dimensional (3D) reflectivity and 3D vertical velocity. Direct radar reflectivity assimilation (e.g., \cite{johnsonComparisonMultiscaleGSIBased2015,wangDirectAssimilationRadar2017,dowellHighResolutionRapidRefresh2022}) requires background ensembles that explicitly include 3D reflectivity to properly constrain storm structures. \cite{wangCosteffectiveHybridEnVar2026} showed that accurate storm analysis depends critically on updating vertical velocity, necessitating its inclusion in the background ensembles. In summary, to address the requirements for convection-allowing multiscale DA, an emulator that provides subhourly predictions over a relatively large domain to represent interactions across at least mesoscale and convective scales while simultaneously predicting the 3D variables critical for convection is needed.

In this study, we develop and evaluate a convection-allowing ML emulator, MAPCast. This emulator is specifically designed to emulate the Model for Prediction Across Scales (MPAS; \cite{skamarockMultiscaleNonhydrostaticAtmospheric2012}) and to generate surrogate background ensembles for estimating convection-allowing BECs. We aim to address the key questions: Can the MAPCast surrogate background ensembles faithfully reproduce the MPAS BECs? How is the MAPCast surrogate BEC performance dependent on the scales? To address the scientific objectives of the study, MAPCast operates at 3-km convection-allowing resolution with a spatial extent and temporal frequency (15min) that allow representing mesoscale and convective scales and their interactions. In addition, MAPCast explicitly predicts 3D reflectivity and vertical velocity along with other atmospheric state variables. Rather than focusing solely on free forecast skill, this study emphasizes the evaluation of ensemble BEC behavior. Specifically, we assess spread characteristics together with univariate and multivariate correlations for the full-scale fields and their scale-dependent statistics. The evaluations are conducted using 10 retrospective convective storm cases, with direct comparisons of covariance structures derived from MAPCast and MPAS background ensembles valid at 15-m and 1-h lead times. The evaluation of the 15-min and 1-h BECs aims to explore potential implementation of MAPCast surrogate background ensembles within subhourly and hourly DA systems like WoFs and HRRR/Rapid Refresh Forecast System (RRFS), respectively. 

The ultimate objective of MAPCast is to provide a reliable surrogate for physics-based background ensembles in supporting multiscale DA systems. This study presents the first, baseline implementation of MAPCast and focuses on evaluating its ability to reproduce the MPAS BECs across spatial scales. In other words, the goal of this first MAPCast effort documented in this paper is not to fully optimize MAPCast but rather to provide a benchmark for assessing the capability of ML-generated, convection-allowing surrogate background ensembles to represent multiscale BECs. The findings of this study help identify directions for future development and improvement of MAPCast toward the eventual multiscale DA applications.

The remainder of this paper is organized as follows. Section 2 describes the convection-allowing ML emulator, including the model architecture and training strategy. Section 3 presents the cases and metrics used for systematic evaluations. Section 4 shows the comparisons of short-range forecast characteristics, ensemble spread, and correlation estimation between MAPCast and MPAS background ensembles. Finally, Section 5 summarizes the main conclusions and discusses directions of further enhancing MPAS.

\section*{Training data and model architecture}
\subsection*{Training dataset} 

\begin{figure}[ht] %s state preferences regarding figure placement here

% use to correct figure counter if necessary
%\renewcommand{\thefigure}{2}

\includegraphics[width=\textwidth]{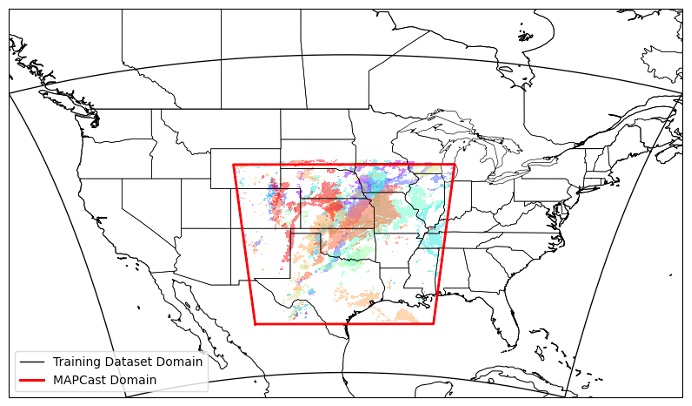}

\caption{\color{Gray} \textbf{The outer domain (black) for the historical MPAS simulation and the inner domain (red) for the MAPCast training. The inner domain covers an 1800 km × 1800 km limited area region with 600 × 600 grid points. The observed composite reflectivity paintballs exceeding 30 dBZ (shaded) from the 10 evaluation cases in Table \ref{tab3} are overlaid.}. }

\label{fig1} % \label works only AFTER \caption within figure environment

\end{figure}

The training dataset is generated using the MPAS model configured in a convection-allowing mode, featuring a horizontal grid spacing of 3 km and 60 vertical levels. The computational domain (black box in Fig.\ref{fig1}) is centered at (38.5°N, 97.5°W) and extends from 25.0° to 47.5°N and from 122.5° to 78.0°W, covering the CONUS. Each MPAS simulation is initialized from operational HRRR analyses. When these analyses are unavailable, Rapid Refresh (RAP) analyses are used. Lateral boundary conditions (LBCs) are provided by the corresponding forecasts. Physical parameterizations (Table \ref{tab1}) are selected to align with the operational Rapid Refresh Forecast System (RRFS) framework (https://github.com/NOAA-GSL/rrfs-workflow/tree/rrfs-mpas-jedi/parm), ensuring both dynamical and physical consistency with current convection-allowing guidance.

\begin{table}[!ht]
%\begin{adjustwidth}{-1.5in}{0in} % comment out/remove adjustwidth environment if table fits in text column.
\centering
\caption{MPAS model physics schemes.}
\begin{tabular}{|l|l|}
\hline
\multicolumn{1}{|l|}{\bf Physics schemes} & \multicolumn{1}{|l|}{\bf Specifications}\\ \hline
Microphysics & Thompson-Eidhammer\cite{thompsonStudyAerosolImpacts2014} \\ \hline
Planetary boundary layer & MYNN\cite{nakanishiDevelopmentImprovedTurbulence2009} \\ \hline
Long- and short-wave radiation & RRTMG\cite{iaconoRadiativeForcingLonglived2008} \\ \hline
Land surface model & RUC\cite{smirnovaModificationsRapidUpdate2016} \\ \hline
\end{tabular}
\label{tab1}
%\end{adjustwidth}
\end{table}

For each initialization time, historical MPAS simulations are integrated to an 18-h lead time, with model outputs archived at a 15-min interval to match the emulator training time step. The training period spans 16 April to 15 June during 2021-2025. To accommodate GPU memory constraints, we restrict the training data to the subdomain indicated by the red box in Fig. 1, covering an area of 1800 km × 1800 km. The extended temporal coverage and large spatial domain sample a broad spectrum of mesoscale environments and convective processes, including isolated supercells, organized squall lines, and mesoscale convective systems. This diversity enables MAPCast to learn representative multiscale interactions and flow-dependent short-term error growth across a range of convective storms.

The state vector used for training includes 3-D atmospheric variables, horizontal wind components (\textit{u} and \textit{v}), potential temperature (\textit{T}), pressure (\textit{p}), and water vapor mixing ratio (\textit{q}), as well as two-dimensional surface and near-surface fields, including 2-m temperature (t2), 2-m water vapor mixing ratio (q2), and 10-m wind components (u10 and v10). To enable MAPCast as an ML-based surrogate background ensemble generator for convection-allowing ensemble DA, additional critical variables, 3D vertical velocity (w) and 3D reflectivity (Z), are included to better represent storm-scale dynamics and microphysical structures. Given the GPU memory constraints, only every third level of the 60 available vertical levels is retained. The native hybrid terrain-following vertical coordinate is preserved, rather than interpolating to pressure levels, to facilitate direct coupling with existing DA frameworks such as the Joint Effort for Data assimilation Integration (JEDI; \cite{guerretteDataAssimilationModel2023}). With this configuration, MAPCast predicts a total of 144 variables. 

\subsection*{MAPCast architecture and training}

The convection-allowing ML emulator, MAPCast, is constructed by adapting GraphCast \cite{lamLearningSkillfulMediumrange2023} to a limited-area setting. Figure \ref{fig2} schematically illustrates the overall architecture, which follows an encoder-processor-decoder framework \cite{keislerForecastingGlobalWeather2022,oskarssonGraphbasedNeuralWeather2023,lamLearningSkillfulMediumrange2023} tailored to represent multiscale atmospheric dynamics. The design is motivated by the following three objectives, 1) to transform model states to a graph representation that enables nonlocal interactions, 2) to propagate information efficiently across scales through hierarchical message passing, and 3) to reconstruct high-resolution gridded outputs while preserving physically meaningful spatial correlations. 

In this architecture, the fundamental computational unit is a GNN layer, composed of nodes and edges. Each layer updates node features by aggregating information from incoming edges and the node itself, followed by nonlinear transformations through a multilayer perceptron (MLP). In this study, each MLP contains a single hidden layer of size 512 with the Swish activation function \cite{ramachandranSearchingActivationFunctions2017} and is followed by a Layer-Norm layer \cite{baLayerNormalization2016}. Additional architectural details of encoder, processor, and decoder networks can be found in \cite{lamLearningSkillfulMediumrange2023} and \cite{oskarssonGraphbasedNeuralWeather2023}. We briefly summarize the key components of MAPCast as follows.

\subsubsection*{a. Multiscale mesh construction}

\begin{figure}[ht] %s state preferences regarding figure placement here

% use to correct figure counter if necessary
%\renewcommand{\thefigure}{2}

\includegraphics[width=\textwidth]{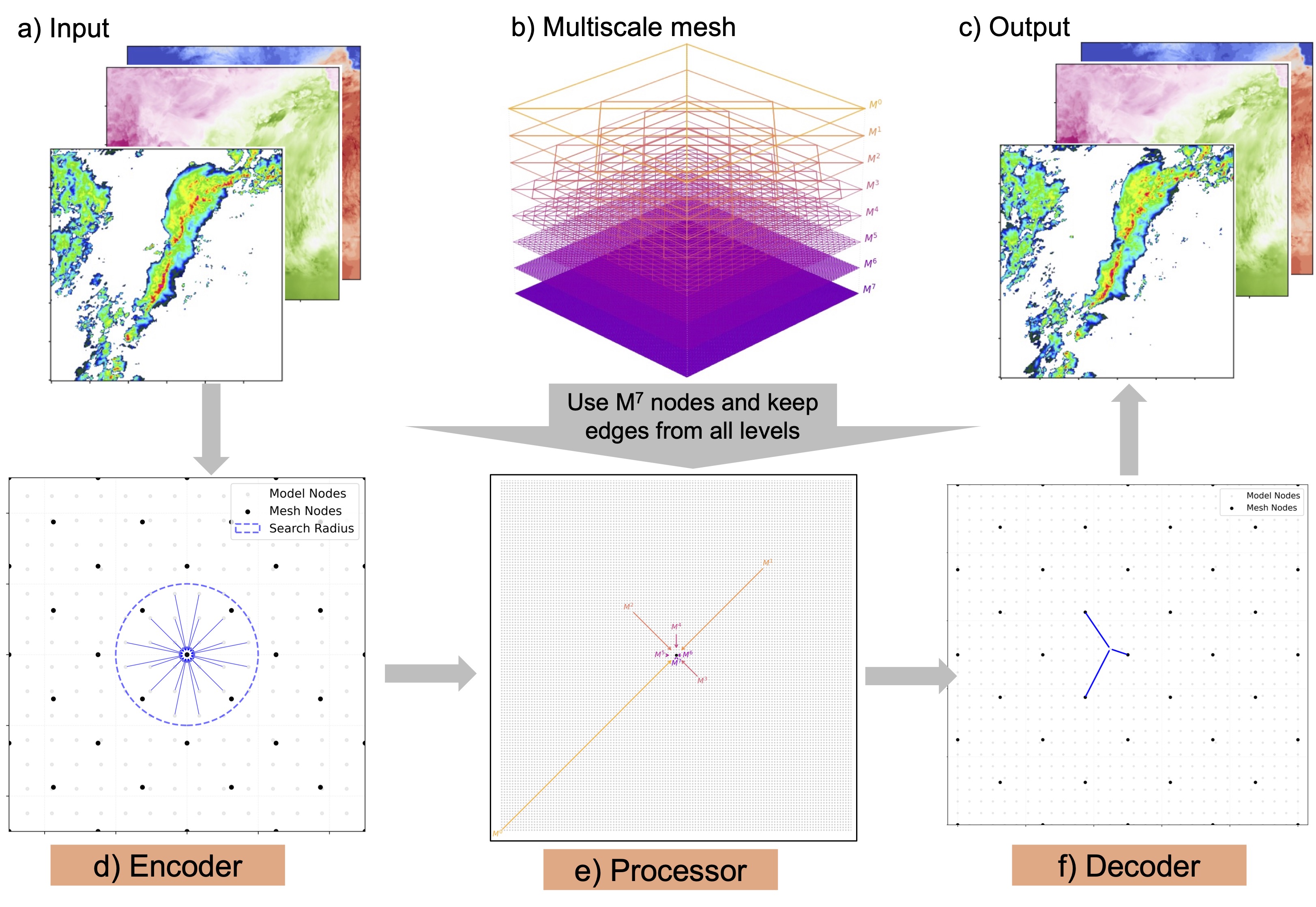}

\caption{\color{Gray} \textbf{Conceptual diagram of the MAPCast architecture. Panels a) and c) show input and output model states, respectively. Panel b) represents the multiscale mesh construction. Panels (d-f) illustrate encoder, processor, and decoder}. }

\label{fig2} % \label works only AFTER \caption within figure environment

\end{figure}

The multiscale mesh provides the structural foundation for hierarchical message passing. Like \cite{lamLearningSkillfulMediumrange2023}, the mesh is created through recursive triangular refinement. As shown in Fig. \ref{fig2}b, the square computational domain is first partitioned into four triangular elements, establishing the coarsest level (M0). Each triangle is then subdivided into four smaller triangles to form the next refinement level (M1). This procedure is repeated to produce a hierarchy spanning levels M0 to M7. The nodes on the finest level (M7) act as the primary computational mesh, while edges from all refinement levels are preserved to enable cross-scale connectivity. Table \ref{tab2} summarizes the number of nodes and edges at each level. 

\begin{table}[!ht]
%\begin{adjustwidth}{-1.5in}{0in} % comment out/remove adjustwidth environment if table fits in text column.
\centering
\caption{Statistics of mesh nodes and edges at each refinement level of the multilevel meshes. Edge counts are shown as doubled values because all mesh connections are represented as bidirectional edges.}
\begin{tabular}{|l|l|l|l|l|l|l|l|l|}
\hline
\multicolumn{1}{|l|}{\bf Mesh level} & \multicolumn{1}{|l|}{\bf 0} & \multicolumn{1}{|l|}{\bf 1} & \multicolumn{1}{|l|}{\bf 2} & \multicolumn{1}{|l|}{\bf 3} & \multicolumn{1}{|l|}{\bf 4} & \multicolumn{1}{|l|}{\bf 5} & \multicolumn{1}{|l|}{\bf 6} & \multicolumn{1}{|l|}{\bf 7}\\ \hline
Node count &5 & 13 & 41 & 145 & 545 & 2113 & 8321 & 33025 \\ \hline
Edge count &16	& 56 & 208 & 800 & 3136	& 12416 & 49408 & 197120 \\ \hline
Total edges &16 & 72 & 280 & 1080 & 4216 & 16632 & 66040 & 263160 \\ \hline
\end{tabular}
\label{tab2}
%\end{adjustwidth}
\end{table}

\subsubsection*{b. Encoder, processor, and decoder networks}

The encoder (Fig. \ref{fig2}d) transforms gridded atmospheric states (Fig. \ref{fig2}a) into mesh node features defined on a multi-level mesh (Fig. \ref{fig2}b). Each mesh node connects to nearby model grid points within a 9-km search radius, yielding a total of 231569 encoder edges that connect grid points to mesh nodes. This grid-to-graph transformation preserves fine-scale structures while reducing dimensionality and allowing irregular connectivity. Such transformation facilitates subsequent nonlocal information exchange.

The processor (Fig. \ref{fig2}e) operates on the merged multiscale mesh and consists of 16 GNN layers. Through repeated message passing across hierarchical edges, the receptive field gradually expands, allowing for the representation of both fine-scale local interactions and broader-scale dynamical coupling. This design provides sufficient nonlinear capacity and enables multiscale information propagation essential for capturing small-scale dynamics.

The decoder (Fig. \ref{fig2}) performs the inverse mapping from mesh nodes back to the model grid (Fig. \ref{fig2}c). For each grid point, the three nearest mesh nodes are identified to estimate the forecast tendency, resulting in approximately 1.44M decoder edges. This local interpolation strategy ensures consistency between the graph representation and the native model grid.

\subsubsection*{c. Training configuration}

MAPCast is trained using a gridpoint-based MSE loss function, summed over all grid points and weather state variables. The level-weighted loss strategy introduced by \cite{lamLearningSkillfulMediumrange2023} is used with a greater weight near the surface. The weights across levels sum to 1 for each of the 3D atmospheric variables. The 2D surface variables are given weights of 1.0. The model predicts one-step tendencies without rollout training. A cosine learning-rate scheduler is employed, with a peak learning rate of 0.001 and linear warm-up during the first 50 epochs. Optimization is performed using the AdamW algorithm \cite{loshchilovDecoupledWeightDecay2019} with weight decay of 0.1 and hyperparameters $\beta_1 = 0.9$, $\beta_2 = 0.95$, and $\epsilon = 10^{-8}$. MAPCast includes $\sim 35.5\text{M}$ trainable parameters.

Consistent with prior studies \cite{lamLearningSkillfulMediumrange2023,oskarssonGraphbasedNeuralWeather2023,floraWoFSCastMachineLearning2025}, input variables are normalized using means and standard deviations computed from the training dataset. The standard deviation of one-step differences is additionally used to normalize the residual target, defined as the increment relative to the input state. During both training and inference, LBCs are provided by MPAS data to isolate emulator errors from uncertainties associated with imperfect boundary forcing, following the experimental design of prior ML models \cite{oskarssonGraphbasedNeuralWeather2023}. MAPCast uses two past weather states to predict the next state \cite{lamLearningSkillfulMediumrange2023}. As discussed in the introduction, the goal of the current study is to provide a benchmark to evaluate the capability of MAPCast in emulating the BECs from MPAS rather than fully optimizing it. As stated in the conclusion section, future efforts will focus on improving multiscale representation of MAPCast surrogate BECs, e.g., by exploring a scale-dependent loss function.

\section*{Data and matrices for evaluating surrogate MAPCast ensemble BECs}
\subsection*{Evaluation data}
This study uses 10 convective events over Southern Great Plains in May 2018 and 2019 (Table \ref{tab3}) to evaluate MAPCast in providing surrogate background ensemble for estimating BECs. These cases feature high-impact convective storms, such as mesoscale convective systems, squall lines, and supercells. The spatial distributions of composite reflectivity exceeding 30 dBZ valid at 0045 UTC of each case are shown in Fig. 1. To construct the MPAS background ensemble for each case, we generate 36-member convection-allowing ensemble forecasts using MPAS with a 90-min forecast lead time and 15-min output frequency. The forecasts are initialized from historical ensemble analyses valid at 0000 UTC. These ensemble analyses were produced by the University of Oklahoma (OU) Multiscale data Assimilation and Predictability (MAP) laboratory for the 2018 and 2019 Hazardous Weather Testbed Spring Forecast Experiment (HWT-SFE; \cite{clarkRealTimeSimulatedForecasting2020}). Please refer to \cite{gasperoniComparisonMethodsSample2020} for the details of generating the OU MAP ensemble analyses. The MPAS model configuration, including computational domain and physical parameterizations, is identical to that used for the training dataset described in Section 2.1, ensuring dynamical consistency between training and testing.

\begin{table}[!ht]
%\begin{adjustwidth}{-1.5in}{0in} % comment out/remove adjustwidth environment if table fits in text column.
\centering
\caption{Retrospective cases used for evaluation in this study, including dominant storm features and the corresponding environmental conditions.}
\begin{tabular}{|l|l|}
\hline
\multicolumn{1}{|l|}{\bf Case date} & \multicolumn{1}{|l|}{\bf Morphology and the environmental conditions} \\ \hline
2 May 2018 & \parbox{10cm}{%
    Discrete supercells across TX, OK, and KS induced and supported by a dryline across the eastern TX Panhandle and an outflow boundary near the OK/KS border, with strong instability, rich moisture, and enhanced low-level shear.
  } \\ \hline
3 May 2018 & \parbox{10cm}{%
    Organized convective lines and embedded supercells moving into eastern OK and southeast KS with upscale growth into squall lines, supported by strong instability, steep midlevel lapse rates, and strengthening low-level shear from an intensifying low-level jet.
 } \\ \hline
 14 May 2018 & \parbox{10cm}{%
    Weakening discrete to clustered storms across southwest KS and the TX Panhandle along a retreating surface dryline, supported by strong instability and sufficient shear but increasingly limited by growing CIN, with a later elevated convection aided by a strengthening low-level jet.
 } \\ \hline
30 May 2018 & \parbox{10cm}{%
    Discrete supercells across northwest OK and the eastern TX Panhandle supported by a dryline and an outflow boundary, with strong instability, rich moisture, and enhanced low-level shear.
 } \\ \hline
1 May 2019 & \parbox{10cm}{%
    Isolated severe thunderstorms across west TX, initially along the dryline and now near a synoptic front from northwest TX into southern OK, supported by sufficient instability, with upscale growth into a squall line along the stalling front.
 } \\ \hline
2 May 2019 & \parbox{10cm}{%
    A slow-moving MCS with bowing segments across northwest TX and south-central OK, supported by an outflow boundary from previous convection and a north-south dryline across south-central OK and TX.
 } \\ \hline
6 May 2019 & \parbox{10cm}{%
    Isolated supercells across west TX along the dryline, supported by strong instability, steep mid-level lapse rates, and sufficient shear with increasing low-level helicity.
 } \\ \hline
8 May 2019 & \parbox{10cm}{%
    A leading discrete supercell and additional storms along a quasi-stationary front from southeast KS into northern OK, supported by moderate instability and sufficient shear.
 } \\ \hline
21 May 2019 & \parbox{10cm}{%
    An MCS with embedded and discrete convection from northeastern OK into southern MO and southeast KS along a northward-lifting warm front, supported by moderate instability and strengthening low-level shear.
 } \\ \hline
27 May 2019 & \parbox{10cm}{%
    A squall line with embedded supercells across the TX Panhandle and southern High Plains supported by a dryline and strong large-scale ascent from an approaching shortwave trough, with strong instability, steep lapse rates, and intense deep-layer flow.
 } \\ \hline
\end{tabular}
\label{tab3}
%\end{adjustwidth}
\end{table}

Because reflectivity is not available at the lead time hour 0, MPAS simulations valid at 0015 and 0030 UTC are used as inputs for MAPCast. The subsequent 15-min and 1-h ensemble forecasts from 0030 UTC, i.e., valid at 0045 and 0130 UTC, respectively, are evaluated in this study. The evaluation of 15-min and 1-h surrogate background ensembles is designed to assess the application of MAPCast to provide surrogate background ensembles for both subhourly and hourly DA systems, represented by WoFs and HRRR, respectively. WoFs primarily focuses on forecasts of localized severe convection, whereas HRRR targets a broader range of convective storms, including both isolated storms and organized mesoscale convective systems. As a result, the relevant background-error structures span different spatial scales. This motivates a scale-dependent evaluation of MAPCast surrogate background ensembles, with emphasis on its ability to reproduce background-error statistics for different scales required for future multiscale DA. These emulated background ensembles are compared with the corresponding MPAS background ensembles to assess the fidelity of MAPCast in emulating background forecasts and estimating BECs across scales. 

%\clearpage makes sure that all above content is printed at this point and does not invade into the upcoming content
%\clearpage

\subsection*{Evaluation matrices}

In this study, the evaluation metrics are computed for each event and forecast lead times of 15 and 60 min and then aggregated over the 10 cases. This aggregation provides both mean performance and robustness across different storm environments. Throughout this study, the MPAS ensemble serves as the reference, and the focus is to evaluate how closely the MAPCast short-term ensemble BECs match those of the MPAS ensemble BECs. 

Ensemble BECs include two key components, ensemble spread and correlations. Ensemble spread serves to estimate the background error uncertainty, essentially measuring how much confidence to place in the model states versus the incoming observations in DA. Spatial and cross-variable correlations determine how to spread observation information across the space and variables. Therefore, the metric suite evaluates the MAPCast ensemble BECs from multiple complementary perspectives, including amplitude (variance ratio) and spatial distribution (spread-pattern correlation) of ensemble spread, correlation structures (matrix similarity and error norms), and multivariate coupling. These matrices are also computed at two decomposed scales to achieve the scale-dependent evaluation of MAPCast surrogate BECs.

\subsubsection*{a. Amplitude and spatial distribution of ensemble spread}

To evaluate whether the surrogate MAPCast background ensemble reproduces BECs estimated from the MPAS background ensemble, we compare ensemble background-error statistics for each event and prognostic field. Let $\mathbf{x}_k^{\text{S}}$ and $\mathbf{x}_k^{\text{M}}$, $k = 1, 2, \dots, K$, denote the MAPCast and MPAS background ensembles, respectively, where $K$ is the ensemble size. Ensemble perturbations are computed relative to their corresponding background ensemble mean following
\begin{align}
\mathbf{x}_k'^{\text{S}} &= \mathbf{x}_k^{\text{S}} - \overline{\mathbf{x}}^{\text{S}}, \\
\mathbf{x}_k'^{\text{M}} &= \mathbf{x}_k^{\text{M}} - \overline{\mathbf{x}}^{\text{M}},
\end{align}
where $\overline{\mathbf{x}}^{\text{S}} = \frac{1}{K}\sum_{k=1}^{K}\mathbf{x}_k^{\text{S}}$ and $\overline{\mathbf{x}}^{\text{M}} = \frac{1}{K}\sum_{k=1}^{K}\mathbf{x}_k^{\text{M}}$.

To explore multiscale behavior, perturbations are further decomposed into large and small scales using a Gaussian low-pass filter $G$,
\begin{equation}
\mathbf{x}_{\text{large}}' = G \mathbf{x}' \quad \text{and} \quad \mathbf{x}_{\text{small}}' = \mathbf{x}' - \mathbf{x}_{\text{large}}',
\end{equation}
where $\mathbf{x}'$ represents $\mathbf{x}_k'^{\text{S}}$ and $\mathbf{x}_k'^{\text{M}}$. All covariance matrices are evaluated for the full-scale field and, separately, for the large- and small-scale fields. In this study, while the large-scale statistics reflect the characteristics associated with the ambient environment, the decomposed small-scale statistics link to the storm-scale variability.

The variance ratio,
\begin{equation}
V_R = \frac{\sigma^{\text{S}}}{\sigma^{\text{M}}} = \sqrt{\frac{\sum_{k=1}^{K} \left( x_k'^{\text{S}} \right)^2}{\sum_{k=1}^{K} \left( x_k'^{\text{M}} \right)^2}}, \label{eq:variance_ratio}
\end{equation}
quantifies the overall amplitude of ensemble spread, where $\sigma^{\text{S}}$ and $\sigma^{\text{M}}$ stand for the ensemble standard deviation for the MAPCast and MPAS background ensembles, respectively. A value close to 1 indicates that the MAPCast background ensemble reproduces the correct spread magnitude of the MPAS background ensemble, whereas values greater than 1 indicate overestimation, and values less than 1 indicate underestimation, compared to the MPAS background ensemble.

To assess whether uncertainty is placed in the correct spatial locations, we compute the spread-pattern correlation between the MAPCast and MPAS spread distributions,
\begin{equation}
r_{\text{spread}} = \text{corr}\left(\sigma_n^{\text{S}}, \sigma_n^{\text{M}}\right) = \frac{\sum_{n=1}^{N} \left( \sigma_n^{\text{S}} - \overline{\sigma}^{\text{S}} \right) \left( \sigma_n^{\text{M}} - \overline{\sigma}^{\text{M}} \right)}{\sqrt{\sum_{n=1}^{N} \left( \sigma_n^{\text{S}} - \overline{\sigma}^{\text{S}} \right)^2} \sqrt{\sum_{n=1}^{N} \left( \sigma_n^{\text{M}} - \overline{\sigma}^{\text{M}} \right)^2}}, \label{eq:spread_corr}
\end{equation}
where $\overline{\sigma}^{\text{S}}$ and $\overline{\sigma}^{\text{M}}$ are the spatial means of $\sigma^{\text{S}}$ and $\sigma^{\text{M}}$, respectively, and $n = 1, 2, \dots, N$, in which $N$ represents the number of model grid points. Higher values indicate that the MAPCast ensemble better locates regions of large and small uncertainty, consistent with the MPAS ensemble. The two metrics at large and small scales are also computed using $x_{\text{large}}'$ and $x_{\text{small}}'$, respectively, to perform the scale-dependent evaluation of MAPCast ensemble spread.

\subsubsection*{b. Ensemble correlation structure and error magnitude}

To directly compare background ensemble correlation structure, we construct sampled correlation matrices from ensemble perturbations. At selected grid points, the normalized perturbation matrix $\mathbf{Z}$, with columns $\mathbf{z}_k = \mathbf{x}_k' / \sqrt{K-1}$ for $k = 1, 2, \dots, K$, yields the sampled covariance matrix 
\begin{equation}
\mathbf{B} = \mathbf{Z} \mathbf{Z}^{\text{T}}.
\end{equation}
In this study, we select 100 grid points randomly for each case. The correlation matrix $\mathbf{C}$—including $\mathbf{C}^{\text{S}}$ for the MAPCast background ensembles and $\mathbf{C}^{\text{M}}$ for the MPAS background ensembles—is obtained by normalizing the covariance matrix with the standard deviations of the respective variables and grid points:
\begin{equation}
\mathbf{C} = \mathbf{D}^{-1} \mathbf{B} \mathbf{D}^{-1}, \label{eq:corr_matrix}
\end{equation}
where $\mathbf{D} = \text{diag}\left(\sqrt{B_{11}}, \sqrt{B_{22}}, \dots, \sqrt{B_{VV}}\right)$ is a diagonal matrix containing the square roots of the diagonal elements of $\mathbf{B}$. Here, $V$ represents the product of the number of grid points and the number of prognostic variables.
Two complementary metrics are used to evaluate the correlation structure and correlation error magnitude. The matrix cosine similarity,
\begin{equation}
\rho_{\text{B}} = \frac{\langle \mathbf{C}^{\text{S}}, \mathbf{C}^{\text{M}} \rangle_{\text{F}}}{\|\mathbf{C}^{\text{S}}\|_{\text{F}} \|\mathbf{C}^{\text{M}}\|_{\text{F}}} = \frac{\text{tr}\left( (\mathbf{C}^{\text{S}})^{\text{T}} \mathbf{C}^{\text{M}} \right)}{\|\mathbf{C}^{\text{S}}\|_{\text{F}} \|\mathbf{C}^{\text{M}}\|_{\text{F}}}, \label{eq:matrix_cosine_similarity}
\end{equation}
measures the structural alignment between MAPCast and MPAS correlation matrices. Values approaching 1 indicate that the two BECs share a similar spatial correlation structure, independent of amplitude. The normalized Frobenius difference
\begin{equation}
D_{\text{B}}=\frac{\|\mathbf{C}^{\text{S}}-\mathbf{C}^{\text{M}}\|_{\text{F}}}{\|\mathbf{C}^{\text{M}}\|_{\text{F}}}
\end{equation}
provides a direct measure of correlation error magnitude. Smaller values indicate closer agreement, reflecting reduced magnitude discrepancy. Together, $\rho_{\text{B}}$ and $D_{\text{F}}$ provide complementary information: the former emphasizes structural similarity, while the latter quantifies absolute error. The two metrics are applied to evaluate both the univariate and multivariate correlations. We also compute these metrics at large and small scales using $\mathbf{x}_{\text{large}}'$ and $\mathbf{x}_{\text{small}}'$, respectively, to perform the scale-dependent evaluation of MAPCast ensemble correlations.

\section*{Evaluation of MAPCast}

In this section, we evaluate MAPCast ensembles mostly using the aggregated matrices across 10 events. Evaluations of the MAPCast ensemble forecasts from the 21 May 2019 MCS case are used as examples to understand the aggregated matrices.

\subsection*{Evaluation of MAPCast short-term individual forecast members}

While the focus of the study is to evaluate how closely the MAPCast short-term background ensemble BECs match those of the MPAS ensemble BECs, we begin with fractions skill score (FSS; \cite{robertsScaleSelectiveVerificationRainfall2008}) and root-mean-square-difference (RMSD) between MAPCast and MPAS ensembles for individual ensemble members to evaluate the skills of MAPCast in reproducing MPAS short-term forecasts. As a first step, neighborhood-based FSS with a neighborhood radius of 48 km is first applied to compare the individual ensemble members in their composite reflectivity prediction skills within the 1-h forecast lead time. The neighborhood radius of 48 km is similar to earlier studies (e.g., \cite{dudaComparingAssimilationRadar2019,johnsonImpactsInitialCondition2024,gasperoniComparisonMethodsSample2020,yangImpactRadarReflectivity2023,wangImprovingCONUSConvectiveScale2024}. The corresponding verification radar observations are obtained from the Multi-Radar Multi-Sensor (MRMS; \cite{smithMultiRadarMultiSensorMRMS2016}). Root-mean-square-difference (RMSD) between MAPCast and MPAS ensembles is also used to show their differences in the predicted continuous fields. 

\begin{figure}[ht] %s state preferences regarding figure placement here

% use to correct figure counter if necessary
%\renewcommand{\thefigure}{3}

\includegraphics[width=\textwidth]{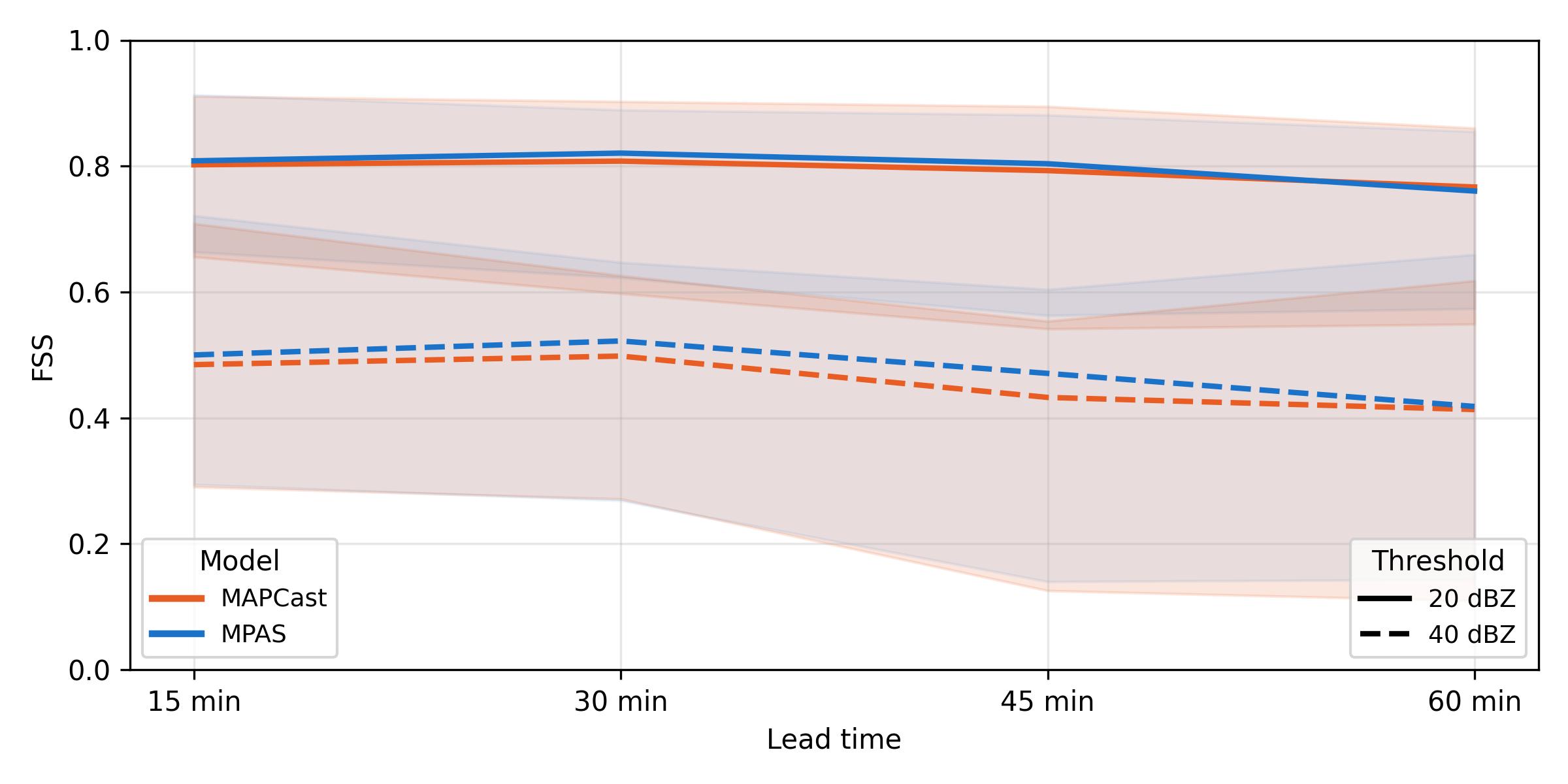}

\caption{\color{Gray} \textbf{Fractions skill scores (FSSs) of composite reflectivity aggregated over 10 retrospective events (lines) for MAPCast and MPAS ensembles at the thresholds of 20 and 40 dBZ. The lower and upper boundaries of the shaded colors indicate the minimum and maximum FSS values across 10 cases. Composite reflectivity represents the vertical-column reflectivity maxima}. }

\label{fig3} % \label works only AFTER \caption within figure environment

\end{figure}

\begin{figure}[!ht] %s state preferences regarding figure placement here

% use to correct figure counter if necessary
%\renewcommand{\thefigure}{4}

\includegraphics[width=\textwidth]{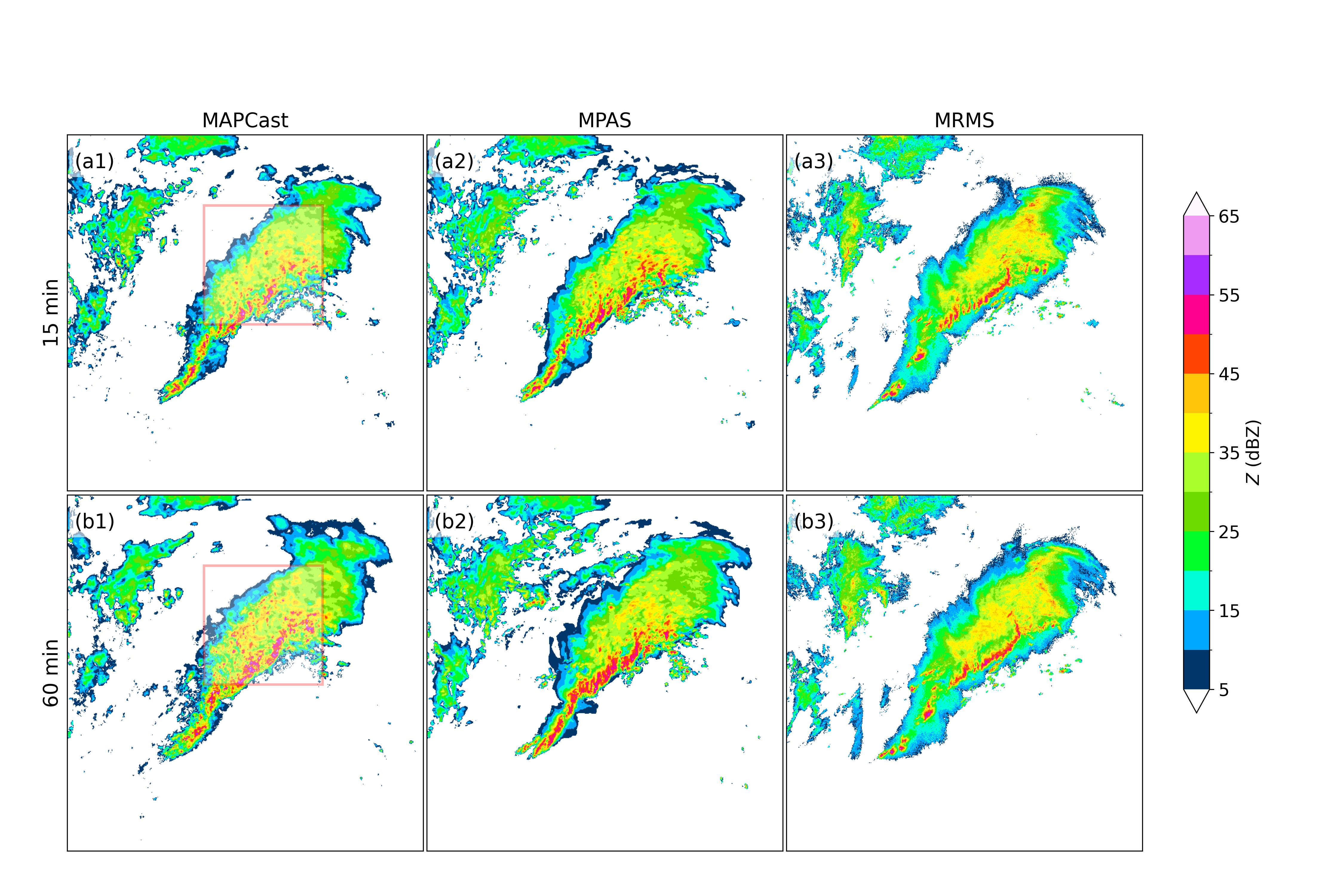}

\caption{\color{Gray} \textbf{The (a1, a2) 15-min and (b1, b2) 60-min composite reflectivity (dBZ) forecasts initialized from 0030 UTC 21 May 2019 from one member of (a1, b1) MAPCast and (a2, b2) MPAS ensembles. The corresponding observed composite reflectivity distributions are included in (a3) and (b3). Red boxes in (a1, b1) denote the area for statistics in Fig. \ref{fig7}}. }

\label{fig4} % \label works only AFTER \caption within figure environment

\end{figure}

Figure \ref{fig3} compares the performance of individual MAPCast and MPAS ensemble members in predicting storm coverage and intensity within a 1-h forecast lead time, verified against MRMS composite reflectivity. At both the 20- and 40-dBZ thresholds, MAPCast presents forecast skill comparable to MPAS, as measured by FSS. Although MAPCast shows a slight degradation (up to ~0.03 lower FSS) at the 40-dBZ threshold compared to MPAS, the overall skill remains similar across all lead times. The shaded regions indicate case-to-case variability and show that the comparable performance between MAPCast and MPAS is consistent across the 10 events. To qualitatively illustrate these results, Fig. \ref{fig4} presents representative forecasts from the 21 May 2019 MCS case at 15- and 60-min lead times. Both MAPCast and MPAS reproduce the overall structure and orientation of the convective system, yielding composite reflectivity patterns that closely align with MRMS observations (Figs. \ref{fig4}a3 and \ref{fig4}b3). The agreement is particularly evident in the spatial extent and organization of the stratiform and convective regions. Notable differences appear in the intensity distribution. Discrepancies between MAPCast and MPAS are primarily confined to reflectivity values greater than 35 dBZ. MAPCast tends to produce weaker convective cores, with reduced peak reflectivity, while exhibiting slightly enhanced reflectivity in the surrounding stratiform regions compared to MPAS. These characteristics are consistent with the aggregated FSS statistics in Fig. \ref{fig3}, where differences are more pronounced at the higher reflectivity threshold. 

\begin{figure}[ht] %s state preferences regarding figure placement here

% use to correct figure counter if necessary
%\renewcommand{\thefigure}{3}

\includegraphics[width=\textwidth]{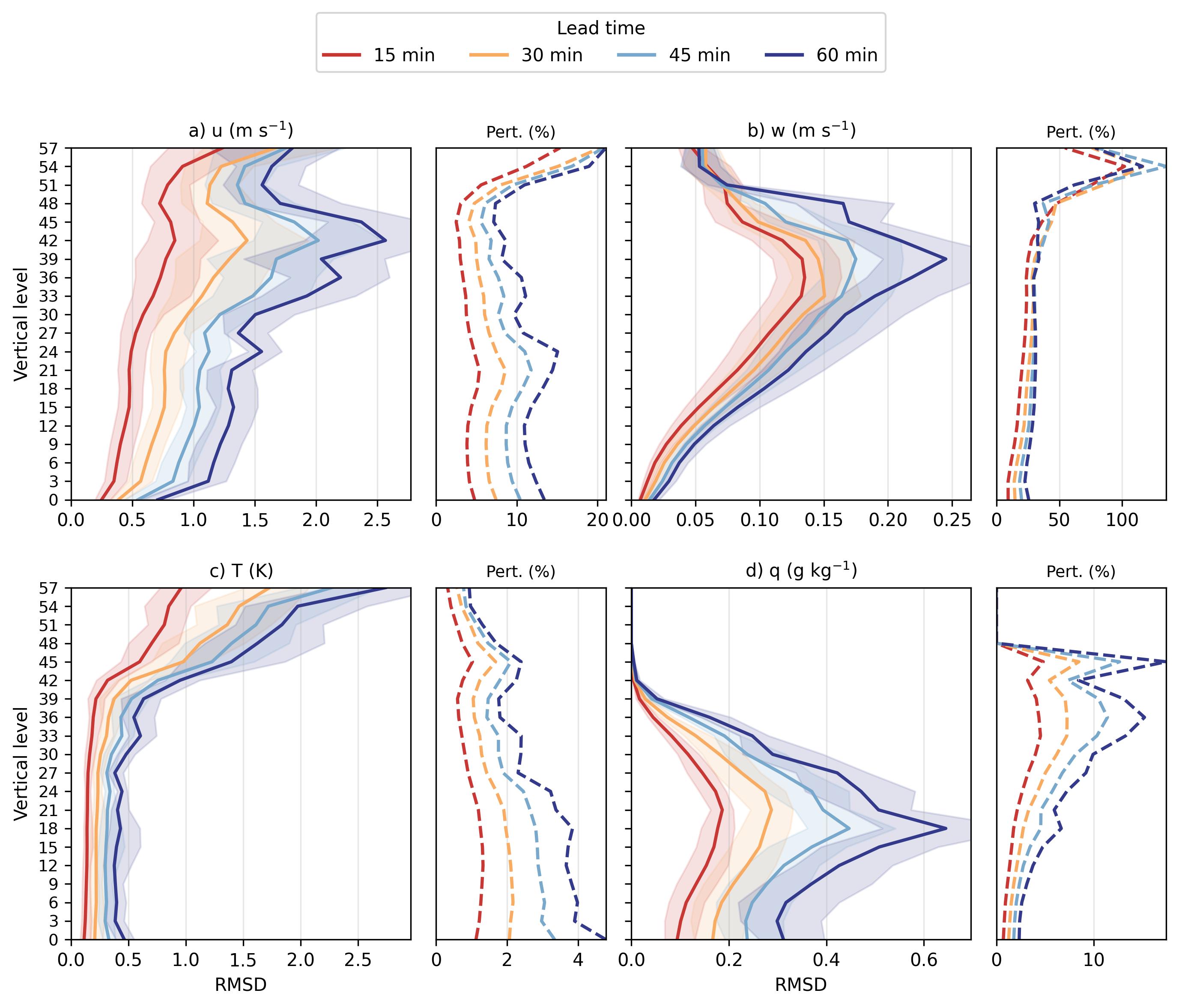}

\caption{\color{Gray} \textbf{Profile of RMSDs (solid lines) and normalized RMSDs (unit: \%; dashed lines) for (a) u, (b) w, (c) T, and (d) q between MAPCast and MPAS ensemble forecasts at 15-, 30-, 45-, and 60-min lead times. The boundaries of the shaded colors indicate the minimum and maximum RMSD values across 10 cases. The normalized RMSDs are computed by dividing RMSDs over the mean of the absolute value of MPAS ensemble forecasts over 10 cases}. }

\label{fig5} % \label works only AFTER \caption within figure environment

\end{figure}

Figure \ref{fig5} presents vertical profiles of the RMSDs between MAPCast and MPAS for the representative 3D variables (u, w, T, and q), computed across ensemble members and aggregated over the 10 events. These variables are chosen as representative fields to characterize differences in dynamical, thermodynamic, and moisture structures. At the 15-min forecast lead time, the RMSD of u increases with height, from less than 0.2 m s-1 near the surface to ~1.3 m s-1 near the model top (Fig. \ref{fig5}a), corresponding to ~4\%-15\% of MPAS forecast values. A similar vertical structure is observed for T, with RMSD values increasing from below 0.1 K near the surface to nearly 1.0 K aloft (Fig. \ref{fig5}c). Despite the increasing RMSD, the temperature differences remain below 2\% of the corresponding MPAS forecast values throughout the vertical levels and below 1\% of MPAS forecast values in the uppermost levels. The larger upper-level differences in u and T are consistent with the vertically weighted loss function described in Section 2.2c, which assigns lower weights to upper-level states during training. In contrast, w and q present distinct vertical structures. Their RMSDs remain small in the lower levels but increase rapidly above around the 15th model level (Figs. \ref{fig5}b, d), coinciding with the altitude range where convective motions are most active. Peak RMSD values reach ~0.13 m s-1 for w and 0.18 g kg-1 for q. The RMSD values for q remain below 5\% of the corresponding MPAS forecast values, whereas those of w take 5\%-20\% of the corresponding MPAS forecast values in the most levels and reach 100\% near the model top. Among the evaluated variables, w exhibits the largest relative differences, indicating that w remains the most challenging field for MAPCast to emulate. This limitation is likely related to the near-zero mean and highly heterogeneous distribution of w, which complicates normalization and training, and therefore represents an important direction for future model development. As forecast lead time increases from 15 to 60 min, RMSD values increase for all variables while maintaining similar vertical profile shapes. This behavior suggests that forecast differences between MAPCast and MPAS grow systematically without fundamentally altering their vertical error characteristics. The case-to-case variability reflected by the shaded regions follows the averaged RMSDs and shows the robustness across the 10 events.

\begin{figure}[ht] %s state preferences regarding figure placement here

% use to correct figure counter if necessary
%\renewcommand{\thefigure}{3}

\includegraphics[width=\textwidth]{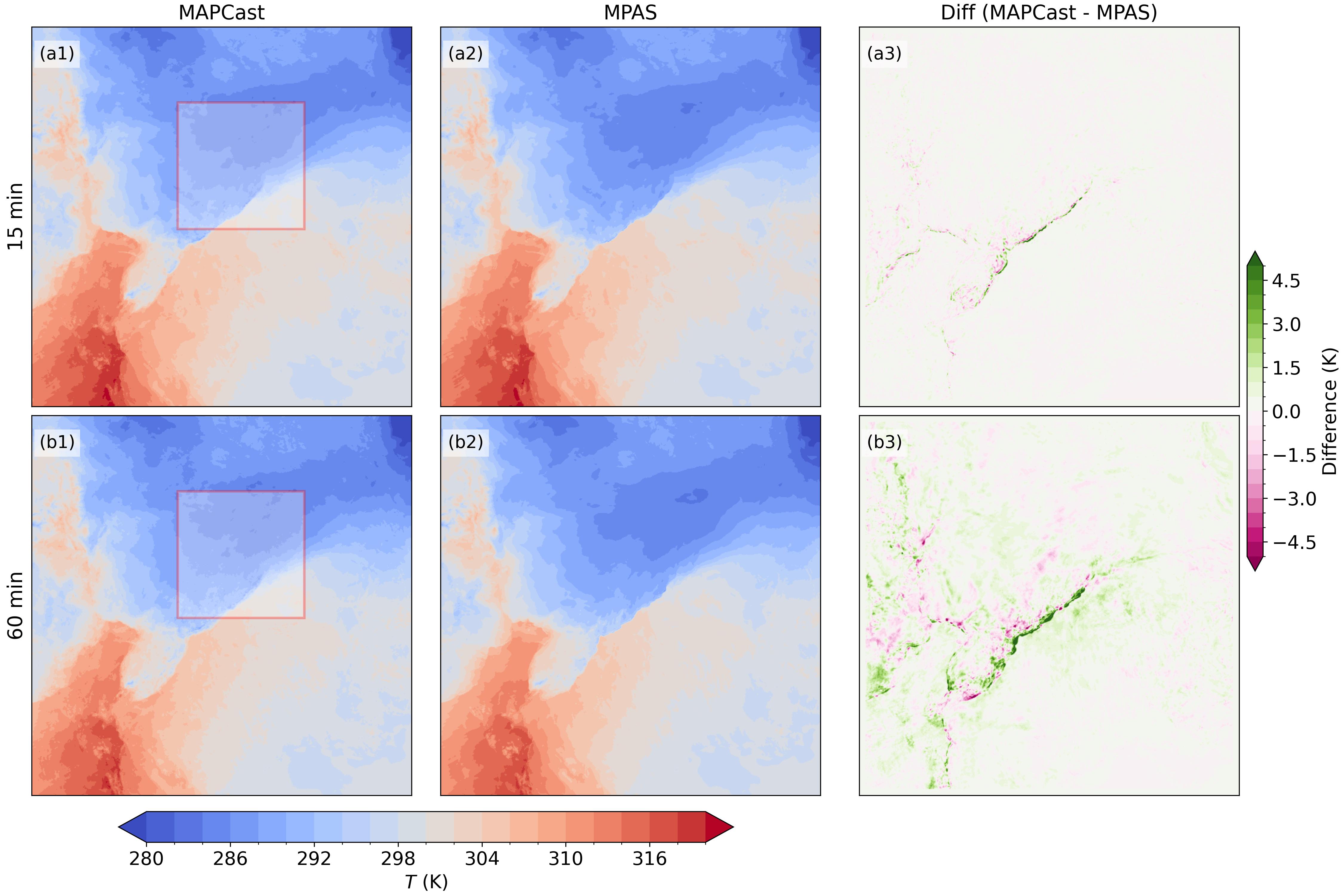}

\caption{\color{Gray} \textbf{The (a1, a2) 15-min and (b1, b2) 60-min T forecasts (K) at the lowest model level initialized from 0030 UTC 21 May 2019 from one member of (a1, b1) MAPCast and (a2, b2) MPAS ensembles. (a3, b3) The rightmost column shows the model differences in T. Red boxes in (a1, b1) denote the area for statistics in Fig. \ref{fig7}}. }

\label{fig6} % \label works only AFTER \caption within figure environment

\end{figure}

To diagnose the T differences shown in Fig. \ref{fig5}, we compare MAPCast and MPAS forecasts in Fig. \ref{fig6}. At the 15-min lead time, MAPCast reproduces the MPAS general structure, including the placement and strength of the cold pools and the southern warm sector (Figs. \ref{fig6}a1, a2). The residual differences between MAPCast and MPAS (Fig. \ref{fig6}a3) are weak and are primarily confined to narrow lines along the cold-pool leading edge. After 45 min, both forecasts provide a comparable southward extension of the cold pools (Figs. \ref{fig6}b1, b2). We also notice that the magnitude and spatial coverage of their differences increase (Fig. \ref{fig6}b3), with the largest differences collocated with regions of strong horizontal T gradients (e.g., cold-pool boundaries). Overall, the discrepancies remain dominated by small-scale, gradient-aligned features, consistent with growing phase errors in sharp boundary positions as forecast lead time increases. 

\begin{figure}[!ht] %s state preferences regarding figure placement here

% use to correct figure counter if necessary
%\renewcommand{\thefigure}{3}

\includegraphics[width=\textwidth]{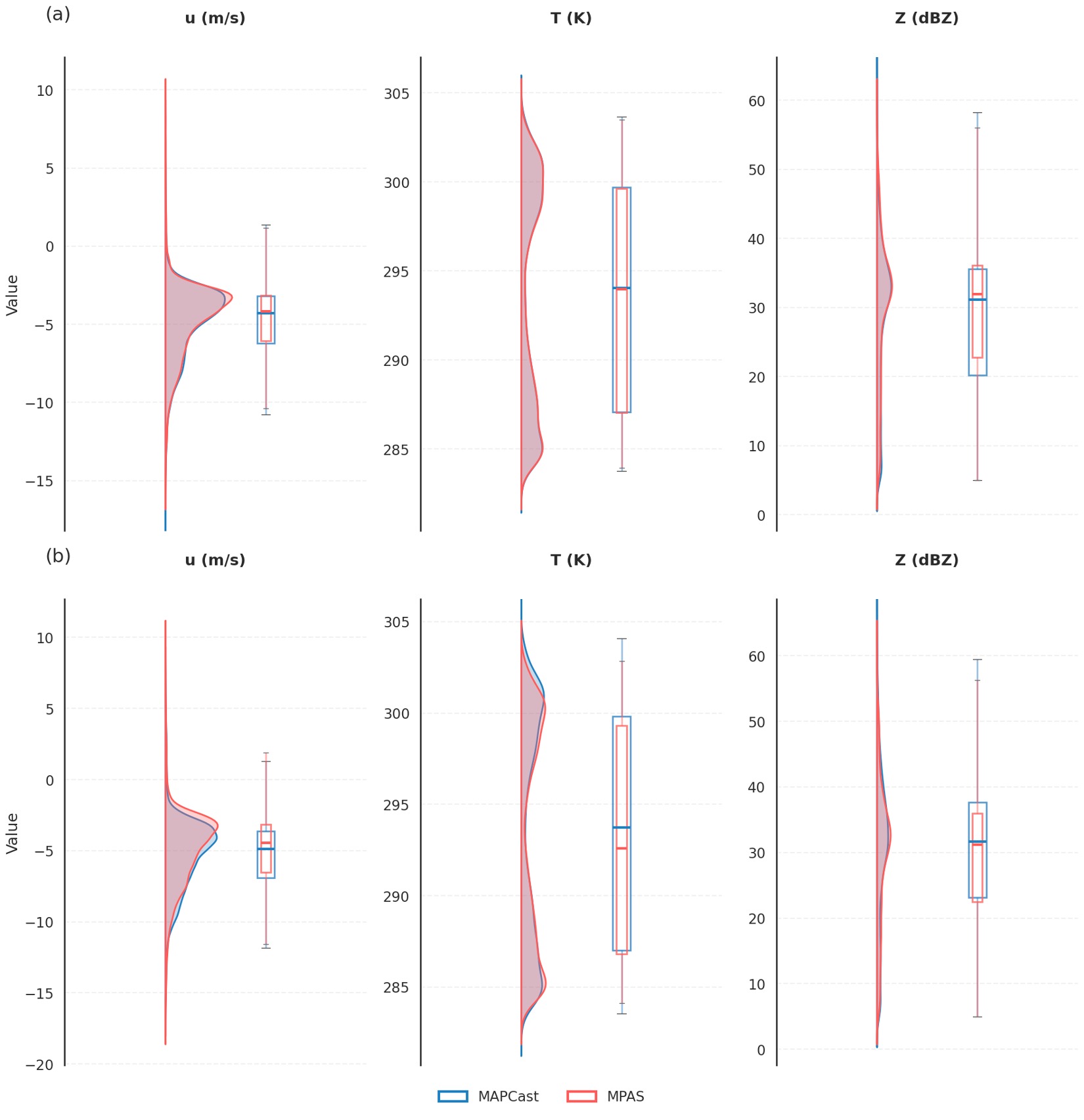}

\caption{\color{Gray} \textbf{Comparisons of data frequency distributions for u, T, and Z at the lowest model level within the red box of Figs. \ref{fig4} and \ref{fig6} between MAPCast and MPAS ensemble forecasts at (a) 15- and (b) 60-min lead times initialized from 0030 UTC 21 May 2019. The red box is chosen to cover the primary feature of storms. Medians are represented by the horizontal lines within rectangles}. }

\label{fig7} % \label works only AFTER \caption within figure environment

\end{figure}

We further quantify the agreement between MAPCast and MPAS individual background ensemble members using the frequency distributions of the predicted u, T, and Z in Fig. \ref{fig7} as examples. At the 15-min lead time, MAPCast closely matches MPAS across all three variables (Fig. \ref{fig7}a). The distributions from MAPCast and MPAS forecasts largely overlap, and the frequency peaks and medians are nearly identical. After 45 min, while the overall distributional shapes remain comparable, the peaks and medians show systematic shifts, most notably for u and T (Fig. \ref{fig7}b). Relative to MPAS, MAPCast displays slightly stronger easterly flow and modest warm shift in the warm sector. In addition, MAPCast shows a slightly higher occurrence at nearly 40 dBZ for Z (Fig. \ref{fig7}b) than MPAS. Such a mismatch is consistent with larger reflectivity differences in the stratiform region (Fig. \ref{fig4}b1) between MAPCast and MPAS.

\begin{figure}[ht] %s state preferences regarding figure placement here

% use to correct figure counter if necessary
%\renewcommand{\thefigure}{3}

\includegraphics[width=\textwidth]{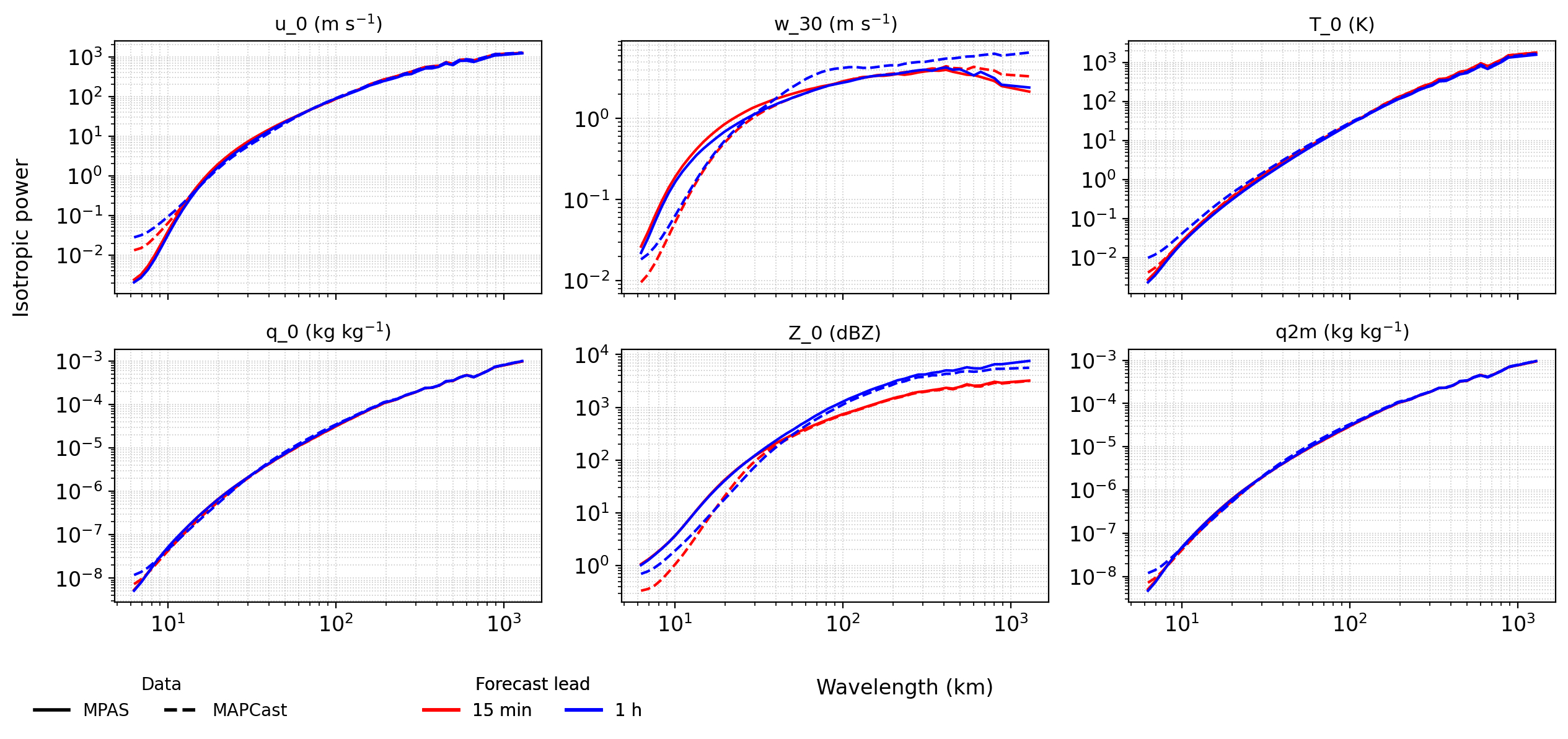}

\caption{\color{Gray} \textbf{Power spectra of selected variables aggregated over 10 cases at (red) 15- and (blue) 60-min lead times for (dashed) MAPCast and (solid) MPAS ensemble forecasts. The selected variables and the corresponding MPAS model level are indicated in the title of each panel, e.g., \textit{u}\_0 stands for the zonal wind at the level 0}. }

\label{fig8} % \label works only AFTER \caption within figure environment

\end{figure}

We further evaluate MAPCast relative to MPAS by comparing spatial power spectra over a broad range of wavenumbers (Fig. \ref{fig8}). At the 15-min lead time, MAPCast and MPAS have nearly identical spectra at relatively large-scale wavelengths for \textit{u}, \textit{T}, \textit{q}, and q2m. The departures are concentrated at the highest wavenumbers, primarily at wavelengths shorter than ~3-4 grid spacings. Among all variables, w and Z show the largest spectral discrepancies, implying that the greatest divergence between the models occurs in storm-related, convective-scale features. At 60 min, MAPCast continues to match MPAS well at large scales for most variables, but the small-scale differences become more pronounced than at 15 min. This result indicates the progressive loss of small-scale phase and amplitude agreement with the increasing lead time. 

The small-scale spectral bias also depends on the variable. MAPCast tends to have higher spectral density than MPAS at small scales for \textit{u}, \textit{T}, and \textit{q}, whereas it exhibits lower small-scale spectral density for w and Z. This contrast is consistent with the spatial characteristics of these fields. Small-scale \textit{w} and \textit{Z} usually feature multiple isolated cells within convection cores, embedded within broad regions of near-zero values (not shown). Additional diagnostics indicate that MAPCast captures the mesoscale organization and placement of convection but underrepresents the intensity and sharp gradients of convective elements, yielding reduced small-scale power in \textit{w} and \textit{Z}. In contrast, small-scale variability in \textit{u}, \textit{T}, and \textit{q} is concentrated within cold pools and their boundaries; MAPCast produces comparatively more disorganized structures in these regions, consistent with the differences in Fig. \ref{fig6}b3, which increases the small-scale variance and thus the spectral densities relative to MPAS.

To sum up, the evaluation of short-term MAPCast individual background ensemble forecasts shows that MAPCast reproduces the overall MPAS forecasts with good fidelity, including realistic representations of storm coverage and temporal evolution, as well as spatial and frequency distributions of state variables and their spectral characteristics. The remaining discrepancies are primarily confined to small scales, particularly in regions with sharp gradients and convective-scale features and in storm-related fields. These results suggest that improving the representation of near-grid-scale structures is the primary priority for further development of MAPCast.

\subsection*{Evaluation of MAPCast background ensemble spread}

As noted in Section 3.2, background ensemble spread provides a flow-dependent estimate of background-error uncertainty. Because finite ensemble forecasts often underestimate true uncertainty, a series of inflation strategies have been proposed to mitigate insufficient spread, such as relaxation to prior perturbations \cite{zhangImpactsInitialEstimate2004}, relaxation to prior spread \cite{whitakerEvaluatingMethodsAccount2012}, additive inflation \cite{dowellAdditiveNoiseStormScale2009}, and multivariate additive inflation tailored for storm-scale DA \cite{wangMultivariateAdditiveInflation2023}. These approaches highlight that, for MAPCast to be viable for ensemble-based DA, its ensembles need to reproduce not only the mean evolution but also the MPAS spread structure and magnitude.

\begin{figure}[!ht] %s state preferences regarding figure placement here

% use to correct figure counter if necessary
%\renewcommand{\thefigure}{3}

\includegraphics[width=\textwidth,height=0.8\textheight,keepaspectratio]{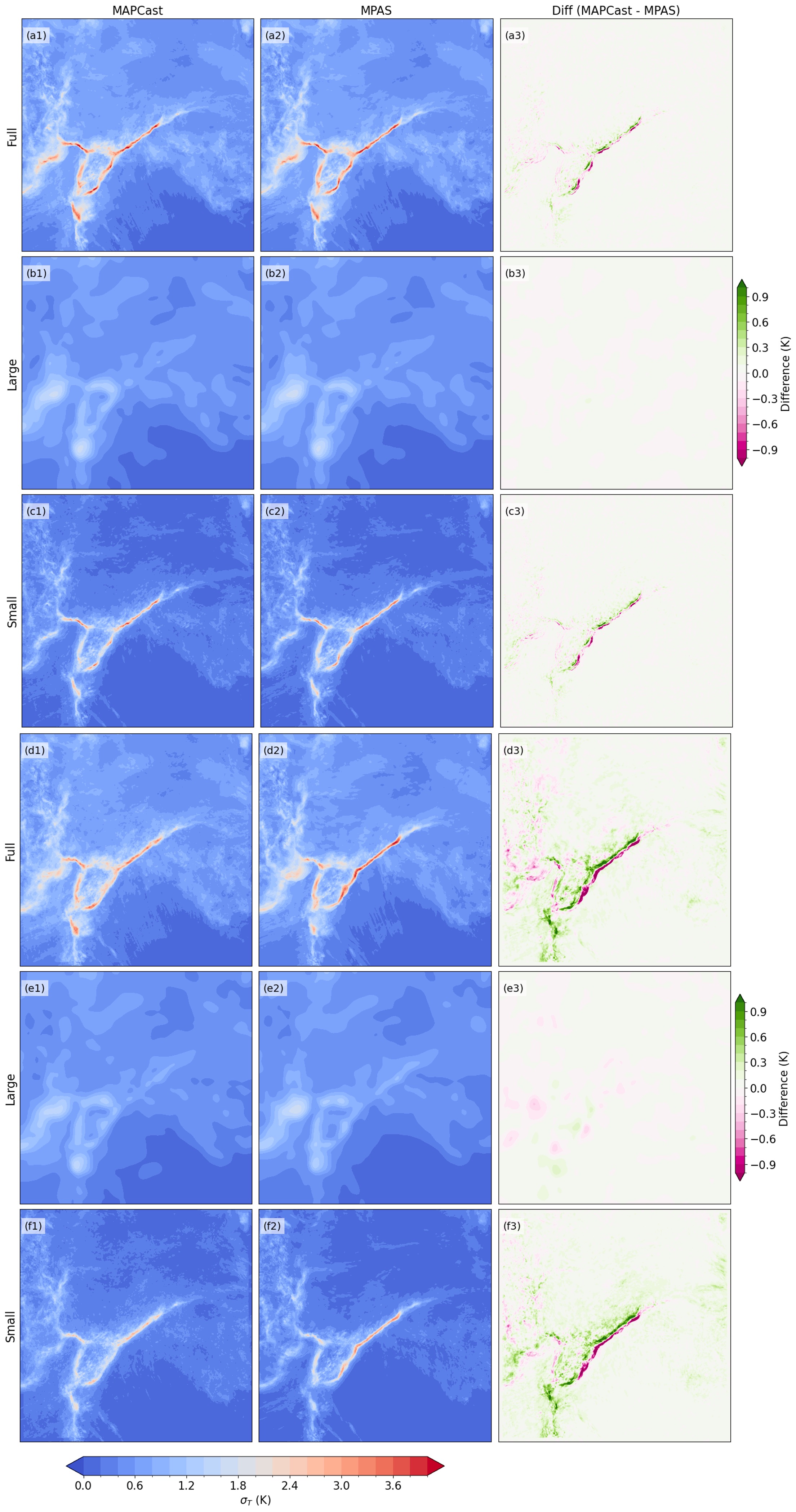}

\caption{\color{Gray} \textbf{Ensemble spread of the lowest model level T from the (a1-a3, b1-b3, c1-c3) 15- and (d1-d3, e1-e3, f1-f3) 60-min (a1-f1) MAPCast and (a2-f2) MPAS ensemble forecasts and (a3-f3) the corresponding model differences. The ensemble forecasts are initialized from 0030 UTC 21 May 2019}. }

\label{fig9} % \label works only AFTER \caption within figure environment

\end{figure}

Figure \ref{fig9} illustrates the comparison for the T ensemble spread at the lowest model level from the 21 May 2019 MCS case. At the 15-min lead time, MAPCast and MPAS ensembles show a highly consistent spread pattern. Strong spread is concentrated near storm-related cold pools, while weak spread is distributed broadly across the surrounding large-scale environment (Figs. \ref{fig9}a1, a2; Figs. \ref{fig6}a1, a2). The residual differences between MAPCast and MPAS ensemble spread (Fig. \ref{fig9}a3) are small and largely confined to the cold-pool leading edge (Fig. \ref{fig9}a3), consistent with the near-boundary differences in Fig. \ref{fig6}a3. To diagnose the physical scale dependence of these spread differences, we decompose ensemble perturbations using a 2-D Gaussian filter with a 30-km length scale and recompute the spread for the large and small scales (Figs. \ref{fig9}b1-b3, \ref{fig9}c1-c3). The large-scale spread captures uncertainty associated with the ambient environment (e.g., the warm sector) and shows negligible differences between MAPCast and MPAS (Fig. \ref{fig9}b3). The small differences imply that MAPCast reproduces MPAS spread well at the large scale. In contrast, the small-scale spread mostly reflects storm-scale variability linked to sharp gradients and convective structures. Although the small-scale spread patterns in MAPCast broadly resemble those in MPAS (Figs. \ref{fig9}c1, c2), the corresponding spread difference (Fig. \ref{fig9}c3) closely mirrors the full-scale spread differences (Fig. \ref{fig9}a3). Therefore, the differences in full-scale spread between MAPCast and MPAS are dominated by small-scale, gradient-aligned features.

At the 60-min lead time, MAPCast and MPAS keep similar overall spread distributions (Figs. \ref{fig9}d1, d2). However, the small-scale spread differences increase relative to 15 min forecasts (Figs. \ref{fig9}f1-f3 vs. Figs. 9d1-d3), while large-scale differences remain small (Figs. \ref{fig9}e1-e3). This scale-selective degradation is consistent with the growing small-scale forecast differences in the state variables (Fig. \ref{fig6}b3). The results of Fig. \ref{fig9} indicate that the primary deficiency of MAPCast spread reproduction lies in representing uncertainty related to sharp, convective-scale boundaries compared to the broader environmental flow.

\begin{figure}[ht] %s state preferences regarding figure placement here

% use to correct figure counter if necessary
%\renewcommand{\thefigure}{3}

\includegraphics[width=\textwidth]{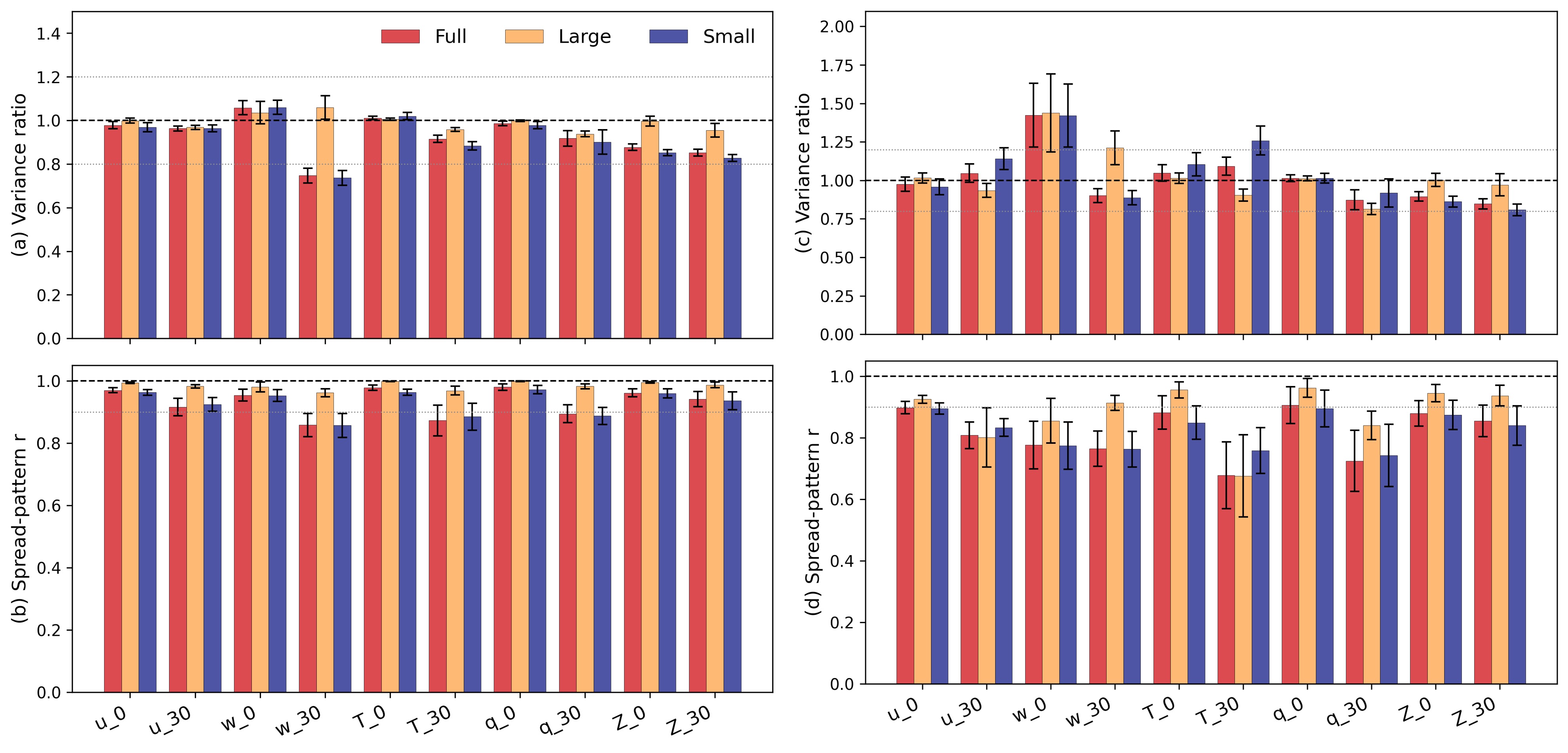}

\caption{\color{Gray} \textbf{(a, c) Variance ratios between MAPCast and MPAS ensemble spread and (b, d) spread-pattern correlations for selected variables at the (\textit{x}\_0) lowest and (\textit{x}\_30) 30th model levels, x stands for u, w, T, q, and Z, aggregated over the 10 evaluated cases. Results are shown for the full (red), large (orange), and small (purple) scales at 15-min (a, b) and 60-min (c, d) forecast lead times. Error bars denote case-to-case variability}. }

\label{fig10} % \label works only AFTER \caption within figure environment

\end{figure}

Figure \ref{fig10} quantitatively assesses the ability of MAPCast ensembles to reproduce MPAS ensemble spread using two complementary diagnostics, the variance ratio between MAPCast and MPAS ensembles and the spread-pattern correlation for all 10 cases. The corresponding results are generally consistent with the findings of Fig. 9. At the 15-min lead time, the full-scale variance ratios for u, T, and q generally exceed 0.9 (Fig. \ref{fig10}a), indicating that MAPCast reproduces the overall spread magnitude for these variables. The best agreement is observed in near-surface fields, with variance ratios approaching 1.0, while upper-level fields show systematically reduced values. For example, the full-scale variance ratio for T is close to 1.0 at the lowest model level but decreases to ~0.9 at the 30th model level. Compared to the conventional variables, the storm-related variables w and Z display larger deviations from 1.0, particularly at higher altitudes. Hence, reproducing the spread amplitude of storm-related variables is more challenging than that of conventional states. The spread-pattern correlations (Fig. \ref{fig10}b) further show that MAPCast accurately reflects the spatial distribution of the ensemble spread from MPAS. Correlations at the lowest model level approach 1.0 for nearly all variables and are consistently higher than those at the 30th model level. 

After scale decomposition, differences among variables become substantially smaller at the large scale. Both the variance ratios and spread-pattern correlations remain close to 1.0 across all variables, indicating that MAPCast captures the magnitude and spatial pattern of the large-scale spread of MPAS remarkably well. In contrast, greater variability among variables appears at the small scale, where the spread statistics largely mirror the behavior of the full-scale results. This consistency indicates that the full-scale discrepancies are primarily influenced by small-scale differences, consistent with Fig. \ref{fig9}. Despite these differences, the case-by-case variability indicated by the error bars in Fig. \ref{fig10} remains within ~0.15 for both metrics, demonstrating robust performance across the 10 events.

At the 60-min lead time, the variance ratios at all scales remain broadly similar to those at 15 min and become slightly larger for most variables (Figs. \ref{fig10}a, c), except for near-surface w, where variance ratios increase by ~40\%-100\%. In contrast, spread-pattern correlations decrease systematically with forecast lead time for all variables and scales (Figs. \ref{fig10}b, d), reflecting progressive decorrelation in spread between MAPCast and MPAS ensembles. Nevertheless, the large-scale spread correlations remain consistently closer to 1.0 than the corresponding full- and small-scale values, indicating that large-scale spread structures are more robustly maintained during the forecast evolution. The case-by-case variability also increases at 60 min, suggesting accumulated forecast divergence at longer lead times. Overall, MAPCast reproduces MPAS ensemble spread more faithfully at shorter lead times and at larger spatial scales.  

\subsection*{Evaluation of MAPCast univariate and multivariate spatial correlations}

Because observations typically sample only a limited subset of variables and spatial locations, univariate and multivariate spatial correlations play a fundamental role in propagating observational information to unobserved model states in ensemble-based DA. Consequently, a viable surrogate background ensemble for DA needs to accurately reproduce the spatial correlation structures of the reference MPAS background ensemble. In practical ensemble-based DA, covariance localization is commonly applied to suppress spurious distant correlations arising from finite ensemble sampling. To dampen the unnecessary, spurious correlations for evaluations, localization is incorporated into the present evaluation following the scale- and variable-dependent localization approach of \cite{wangSimultaneousMultiscaleData2023,wangImprovingCONUSConvectiveScale2024}. Specifically, localization radii are prescribed as functions of both variable type and spatial scale. For the small-scale covariances, all variables use a horizontal cutoff radius of 18 km. For the large-scale correlations, the corresponding localization radii are increased to 60 km for convection-related variables (\textit{w} and \textit{Z}) and 300 km for conventional variables (\textit{u}, \textit{v}, \textit{T}, and \textit{q}). 

\begin{figure}[!ht] %s state preferences regarding figure placement here

% use to correct figure counter if necessary
%\renewcommand{\thefigure}{3}

\includegraphics[width=\textwidth]{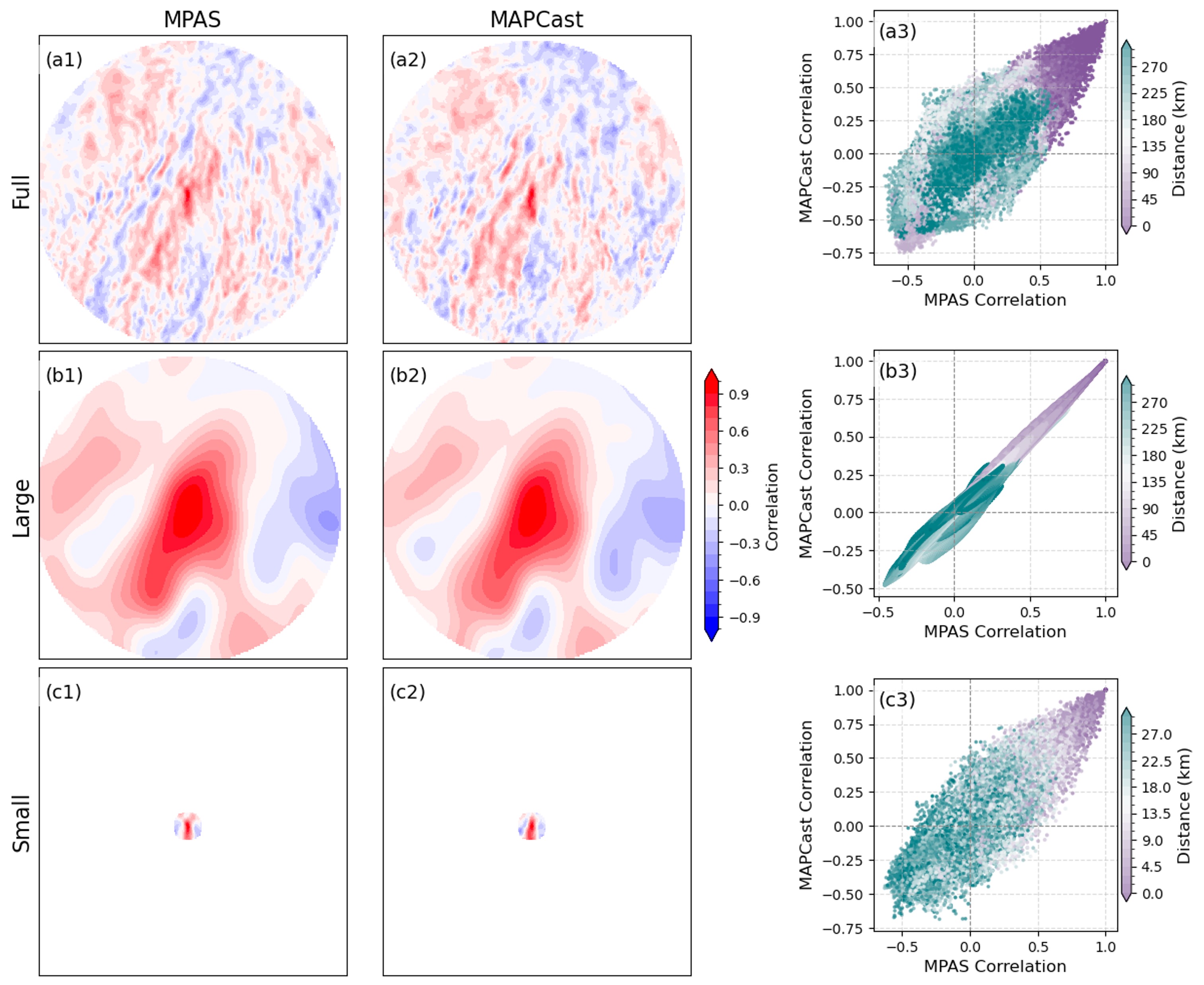}

\caption{\color{Gray} \textbf{The spatial correlation distributions for u at the lowest model level between the point (350, 400) in storms and the entire model grid points constrained using localization radii of 300 km, 300 km, and 18 km for (a1, a2) full, (b1, b2) large, and (c1, c2) small scales, respectively, from (a2-c2) MAPCast and (a1-c1) MPAS ensembles. The spatial u correlation pairs are also shown in a3-c3 for full, large, and small scales. The correlation pairs are formed by computing correlations between each of random 25 points within storms and the mode grid points within the localization radii. The correlations are computed using the 15-min ensemble forecast}. }

\label{fig11} % \label works only AFTER \caption within figure environment

\end{figure}

The agreement between MAPCast and MPAS in estimating spatial correlation for u at the 15-min lead time from the 21 May 2019 MCS case is first assessed in Fig. \ref{fig11}. For the full-scale perturbations (Figs. \ref{fig11}a1-a3), both ensembles show anisotropic correlation patterns, indicating that MAPCast captures the dominant flow-dependent covariance orientation. The scatter distribution in Fig. \ref{fig11}a3 shows strong agreement between MAPCast and MPAS correlations, particularly for great correlations and short distances from the reference point. The spread of points increases as correlations weaken and distance increases, indicating larger discrepancies in weaker distant correlations. The agreement improves further for the large-scale correlation (Figs. \ref{fig11}b1-b3). MAPCast closely reproduces both the magnitude and spatial pattern of the MPAS large-scale correlations, including the broad positive correlation region surrounding the reference point and the weak negative correlations farther away. The scatter plot showcases a compact distribution tightly aligned along the 45° line, demonstrating that MAPCast accurately represents the dominant large-scale correlation structure from MPAS. Hence, MAPCast effectively captures the large-scale background-error pattern that govern the propagation of observational information beyond convective scales in ensemble DA. While MAPCast reproduces the localized structure surrounding the reference point at the small scale, the correlation orientations show differences from those in MPAS (Figs. \ref{fig11}c1, c2). The scatter plot (Fig. \ref{fig11}c3) shows smaller dispersion than for the full scale, but larger dispersion away from the 45° line than that for the large scale. In other words, Figs. \ref{fig11}c1-c3 show the reduced fidelity of MAPCast ensembles in representing convective-scale correlation structures. The degradation at the small scale is consistent with earlier evaluations of spread and spectral characteristics in Sections 4.1 and 4.2. Overall, Fig. \ref{fig11} demonstrates that MAPCast reproduces the spatial correlation structure of MPAS most accurately at the large scale, and differences between MAPCast and MPAS increase at the small scale. Despite that, their differences at both large and small scales are lower than those at the full scale. 

\begin{figure}[ht] %s state preferences regarding figure placement here

% use to correct figure counter if necessary
%\renewcommand{\thefigure}{3}

\includegraphics[width=\textwidth]{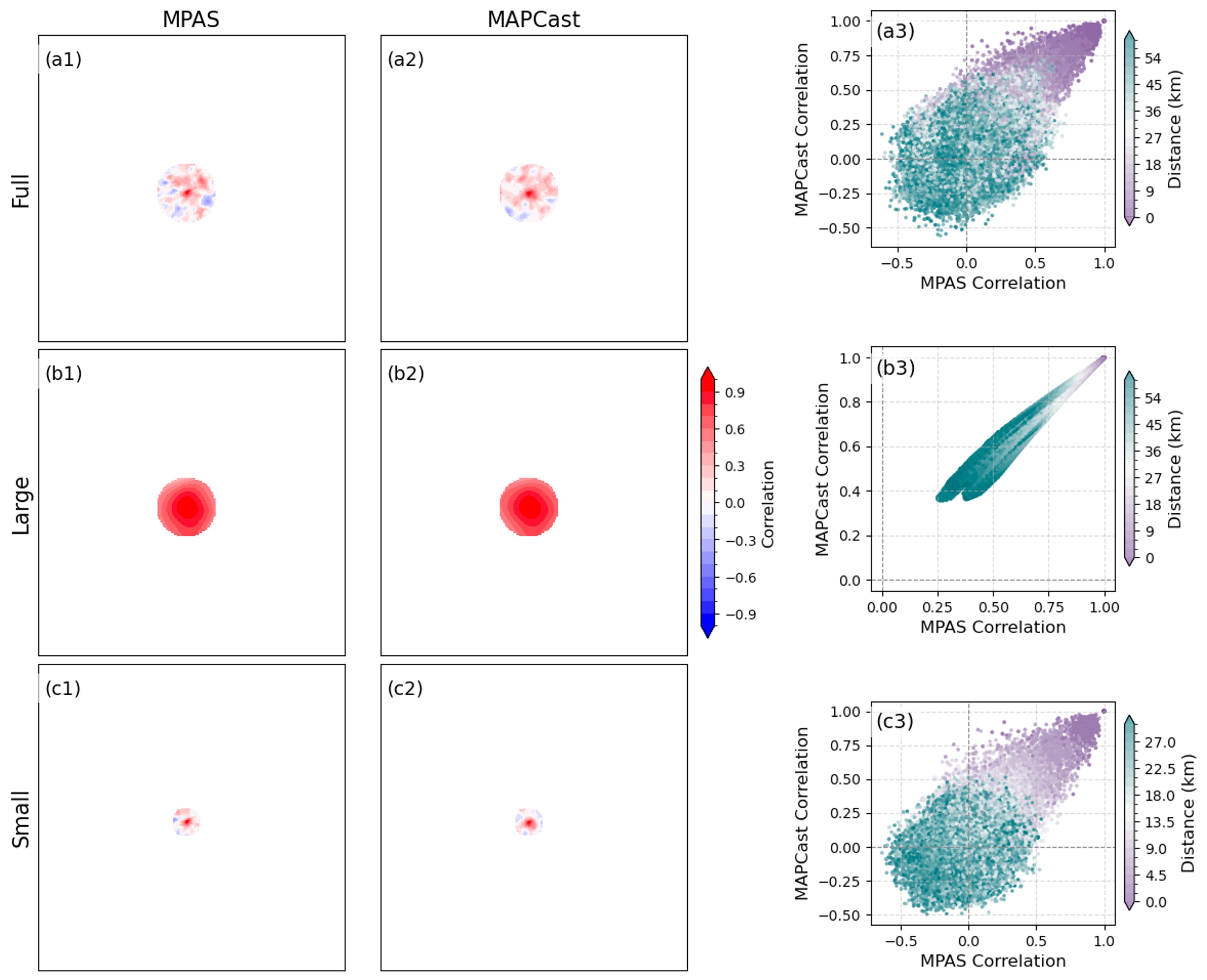}

\caption{\color{Gray} \textbf{Same as Fig. \ref{fig11}, except for Z at the lowest model level with the localization radii of 60 km, 60 km, and 18 km for (a1, a2) full, (b1, b2) large, and (c1, c2) small scales, respectively}. }

\label{fig12} % \label works only AFTER \caption within figure environment

\end{figure}

We further evaluate the agreement between MAPCast and MPAS in estimating spatial correlation for the storm-related variable, \textit{Z}, using the same case in Fig. \ref{fig12}. Compared with the \textit{u} correlations in Fig. \ref{fig11}, the Z correlations are substantially more localized, reflecting the intermittent and highly convective nature of Z-related background errors \cite{wangSimultaneousMultiscaleData2023}. For the full-scale perturbations (Figs. \ref{fig12}a1-a3), both MPAS and MAPCast display compact correlation structures concentrated near the reference point, with alternating positive and negative correlations confined to short distances. MAPCast reproduces the overall spatial extent and orientation of the MPAS correlations. However, the scatter plot in Fig. \ref{fig12}a3 shows larger dispersion than that for \textit{u}, particularly for weak and negative correlations. These results indicate greater uncertainty in representing \textit{Z} correlation structures. At the decomposed large scale (Figs. \ref{fig12}b1-b3), the agreement between MAPCast and MPAS becomes substantially stronger. Both ensembles produce nearly isotropic and smoothly decaying positive correlations centered on the reference point. The corresponding scatter plot forms a narrow distribution tightly aligned with the 45° line, demonstrating that MAPCast accurately reproduces the dominant large-scale \textit{Z} covariance structure. Therefore, MAPCast effectively captures the broader mesoscale organization associated with stratiform and large-scale precipitation structures. For the small-scale correlations (Figs. \ref{fig12}c1-c3), while MAPCast shows weaker, less coherent, and slightly noisier structures than MPAS, MAPCast still well captures the localized nature of the convective-scale correlations. The scatter plot shows better agreements for the small-scale correlations than for its full-scale counterpart (Figs. \ref{fig12}a3 vs c3). Despite that, the scatters substantial spread away from the 45° line, especially for negative correlations and weak amplitudes. These differences indicate that MAPCast has greater difficulty reproducing highly localized \textit{Z} correlations probably associated with intermittent storm-scale variability. Compared with the \textit{u} correlations in Fig. \ref{fig11}, the degradation from large to small scales is more pronounced for \textit{Z}. This behavior is consistent with earlier evaluations showing that storm-related variables, particularly \textit{Z} and \textit{w}, have larger differences in spectra (Fig. \ref{fig8}) and spread (Fig. \ref{fig10}) than conventional variables. Overall, Fig. \ref{fig12} demonstrates that MAPCast reproduces the large-scale spatial correlation structure of \textit{Z} with the highest fidelity, followed by the convective-scale structure. Consequently, the scale-dependent correlation representation is consistently more accurate than that obtained from the full-scale correlations.

\begin{figure}[!ht] %s state preferences regarding figure placement here

% use to correct figure counter if necessary
%\renewcommand{\thefigure}{3}

\includegraphics[width=\textwidth]{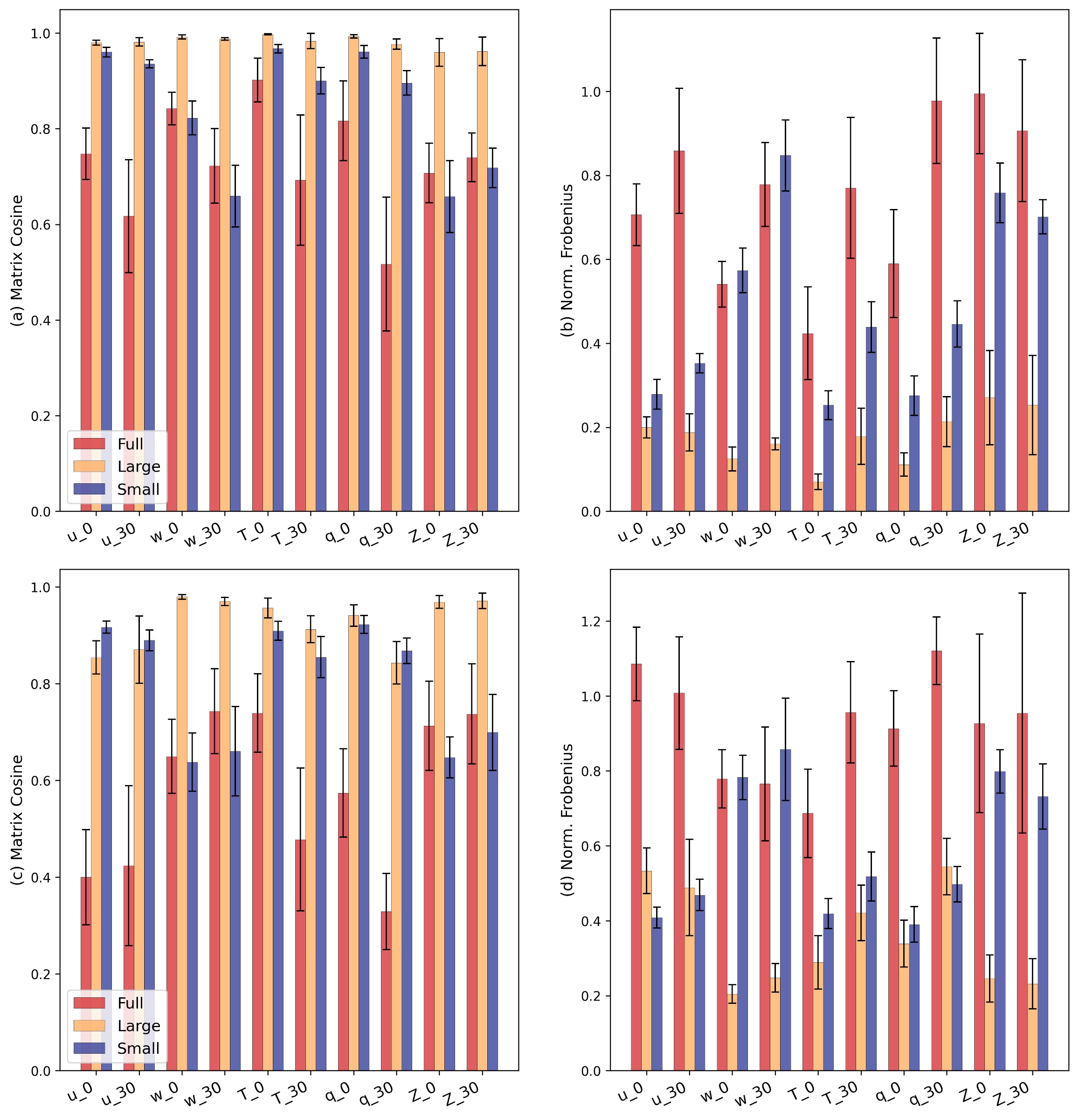}

\caption{\color{Gray} \textbf{(a, c) Matrix cosine similarity and (b, d) normalized Frobenius difference between MAPCast and MPAS ensembles for selected variables aggregated over 10 cases valid at (a, b) 15- and (c, d) 60-min forecast lead times. Both matrices are computed for (red) full, (orange) large, and (blue) small scales. Error bars indicate inter-case variability}. }

\label{fig13} % \label works only AFTER \caption within figure environment

\end{figure}

Figure \ref{fig13} quantitatively assesses the scale-dependent ability of MAPCast ensembles to reproduce MPAS ensemble univariate spatial correlation for all 10 cases using two complementary diagnostics, the matrix cosine similarity and the normalized Frobenius difference. Figure \ref{fig13} generally reflects the findings of Figs. \ref{fig11}, \ref{fig12}. At the 15-min lead time, the large-scale correlations consistently show cosine similarities closest to 1.0 and the smallest normalized Frobenius difference for all selected variables, compared to the full- and small-scale correlations. The case-by-case variability is smallest at the large scale, followed by that at the small scale, with the largest variability at the full scale. This result indicates that MAPCast reproduces the large-scale MPAS univariate spatial correlations with the greatest fidelity and robustness. In contrast, agreement degrades with larger inter-case variability at the full and small scales, reflecting the increasing structural displacement and amplitude mismatch in univariate spatial correlations. For conventional variables, the small-scale correlations are represented more accurately than the corresponding full-scale correlations, as indicated by higher cosine similarities and lower normalized Frobenius differences. Exceptions occur for \textit{w} and \textit{Z} in cosine similarity and for \textit{w} in normalized Frobenius difference. The improved small-scale performance for conventional variables is primarily attributable to the application of stricter localization radii, which suppress weak and noisy distant correlations. In contrast, the remaining degradation for the storm-related variables suggests that the localization scales currently applied to \textit{w} and \textit{Z} may still remain too broad to fully constrain convective-scale correlation structures.

At the 60-min lead time, the large-scale correlations keep the strongest agreement between MAPCast and MPAS ensembles for most variables, with the highest cosine similarities and lowest normalized Frobenius differences among all scales. The only exception is \textit{u}, for which the small-scale correlations are reproduced slightly better than the large-scale correlations. Differences at the full and small scales persist and become more pronounced for most variables, such as \textit{T} and \textit{q}. Relative to the 15-min forecasts, the 60-min MAPCast ensembles show systematically reduced skill in reproducing MPAS univariate spatial correlations, consistent with the increased forecast divergence at longer lead times. 

\begin{figure}[ht] %s state preferences regarding figure placement here

% use to correct figure counter if necessary
%\renewcommand{\thefigure}{3}

\includegraphics[width=\textwidth]{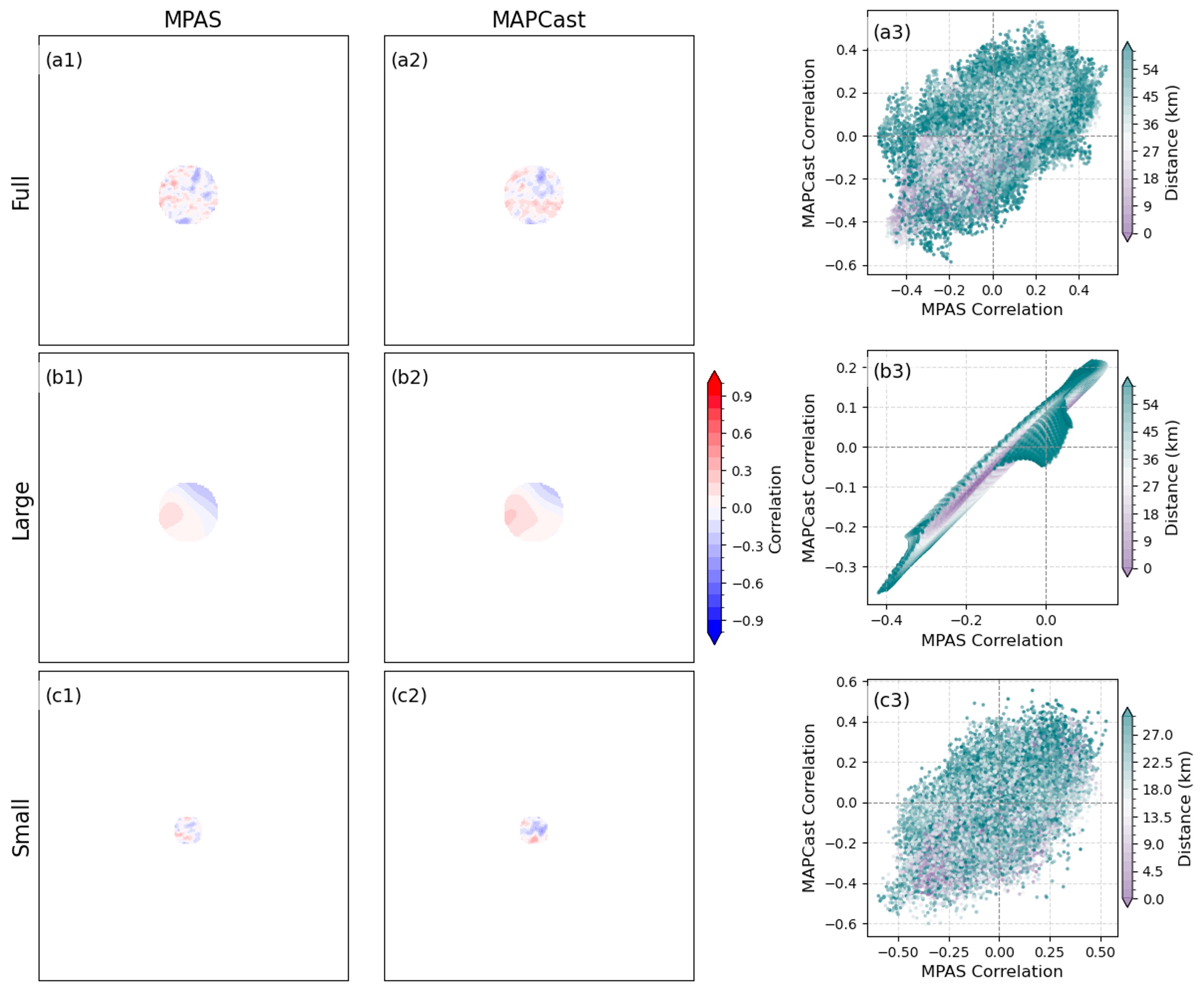}

\caption{\color{Gray} \textbf{Same as Fig. \ref{fig11}, except for correlations between T and Z at the lowest model level}. }

\label{fig14} % \label works only AFTER \caption within figure environment

\end{figure}

Figure \ref{fig14} further assesses the agreement between MAPCast and MPAS in estimating inter-variable spatial correlations between \textit{T} and \textit{Z} at the lowest model level at the 15-min lead time using the 21 May 2019 MCS case. Compared to the univariate spatial correlations (Figs. 11, 12), the \textit{T}-\textit{Z} correlations are weaker, noisier, and more spatially heterogeneous. Such correlation patterns reflect the greater complexity of multivariate coupling than univariate correlations in convection-allowing ensembles. For the full-scale correlation (Figs. \ref{fig14}a1-a3), both background ensembles present localized scatter-like correlation structures near the reference point, with positive and negative correlations distributed asymmetrically within the localization radius. MAPCast fails to reproduce the general orientation and spatial extent of these structures, compared to MPAS. The scatter plot in Fig. \ref{fig14}a3 shows substantial dispersion away from the 45° line, showing large variability in the representation of cross-variable correlations. At the large scale (Figs. \ref{fig14}b1-b3), the agreement between MAPCast and MPAS correlations improves substantially. Both ensembles obtain smooth and spatially coherent dipole structures with similar orientation and amplitude. The corresponding scatter plot is tightly clustered along a line but tilted with the 45° line. This result suggests that MAPCast reproduces the dominant large-scale coupling between \textit{T} and \textit{Z} but with certain biases. Although the small-scale correlations (Figs. \ref{fig14}c1-c3) show slightly a better agreement than the full-scale correlations, but they have much larger discrepancies than the large-scale correlations. The scatter plot displays broad dispersion across both positive and negative correlations, indicating weaker consistency in representing localized multivariate correlation structures. Compared with the corresponding univariate correlations, the degradation in inter-variable spatial correlations shows that reproducing cross-variable coupling is still encouraging, but substantially more challenging than reproducing single-variable correlation structures. 

\begin{figure}[!ht] %s state preferences regarding figure placement here

% use to correct figure counter if necessary
%\renewcommand{\thefigure}{3}

\includegraphics[width=\textwidth]{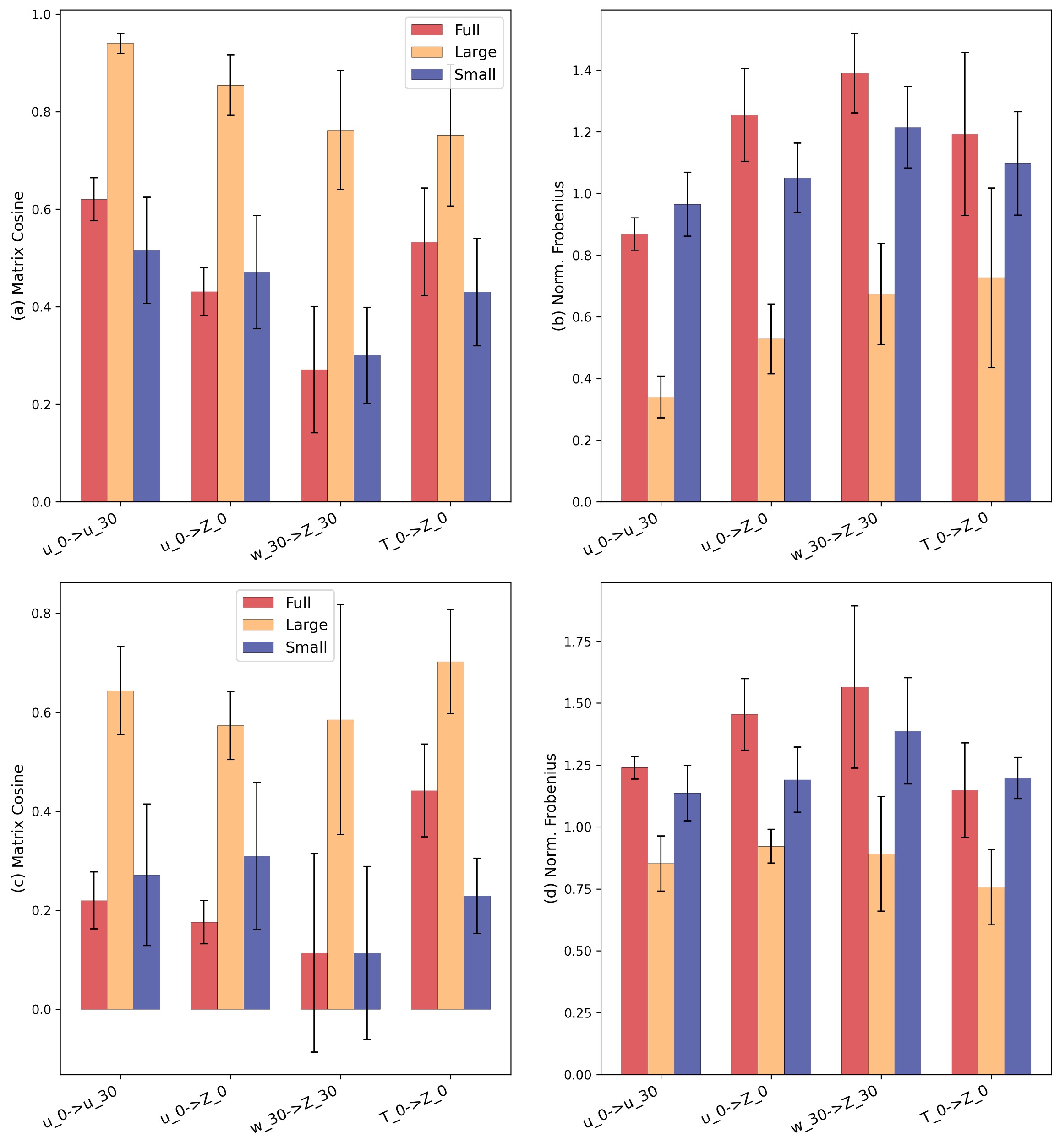}

\caption{\color{Gray} \textbf{Same as Fig. \ref{fig13}, except for cross-variable correlations}. }

\label{fig15} % \label works only AFTER \caption within figure environment

\end{figure}

Figure \ref{fig15} quantitively summarizes the scale-dependent performance of MAPCast in reproducing MPAS cross-variable spatial correlations for all 10 cases using matrix cosine similarity and normalized Frobenius difference for selected variable pairs. Compared with the univariate correlation statistics in Fig. \ref{fig13}, the multivariate metrics present systematically lower cosine similarities and larger Frobenius differences with larger case-by-case variability. These degraded results indicate that reproducing cross-variable covariance structures is more challenging than reproducing single-variable correlations, consistent with Fig. \ref{fig14}. At the 15-min lead time (Figs. \ref{fig15}a, b), the large-scale correlations consistently show the highest cosine similarities and the smallest normalized Frobenius differences across all selected variable pairs. This behavior indicates that MAPCast most accurately reproduces the broad multivariate coupling structures associated with the dynamics beyond convective scales. In contrast, the small-scale correlations manifest substantially weaker agreement. These statistics imply that small-scale multivariate relationships are more sensitive to structural displacement and amplitude errors than their large-scale counterparts. For most variable pairs, the small-scale correlations are reproduced similarly to the full-scale correlations, as reflected by similar cosine similarities and Frobenius differences. The poorer agreement at the small scale than that at the large scale indicates persistent limitations in representing localized small-scale coupling. 

At the 60-min lead time (Figs. \ref{fig15}c, d), the degradation of multivariate correlation structure becomes more evident compared to that at the 15-min lead time (Figs. \ref{fig15}a, b). Cosine similarities decrease and normalized Frobenius differences increase across nearly all scales and variable pairs, accompanied by larger case-to-case variability. Despite this degradation, the large-scale correlations continue to display the strongest agreement between MAPCast and MPAS ensembles. Variable pairs involving storm-related quantities, \textit{w} and \textit{Z}, have the largest degradation, highlighting the difficulty of maintaining dynamically consistent convective-scale coupling during forecast evolution. The weak performance for storm-related variable pairs and at longer lead times indicates that accurate representation of convective-scale cross-variable coupling remains a key requirement for future development of MAPCast to provide surrogate background ensembles for estimating BECs in the context of convection-allowing DA.

\section*{Summary and discussion}

Recent studies have demonstrated that machine learning (ML) can provide computationally efficient emulations of numerical weather prediction (NWP) forecasts across global and regional convection-allowing applications. These advances motivate the exploration of ML emulators for cost-effectively generating short-term surrogate background ensembles to estimate flow-dependent background-error covariances (BECs) in ensemble-based data assimilation (DA). Previous studies have only begun to examine the use of surrogate background ensembles within global coarse-resolution ensemble DA frameworks. Investigations of ML-generated convection-allowing surrogate background ensembles for BEC estimation remain particularly limited. To identify the capability of ML emulators in providing surrogate background ensembles for estimating convection-allowing BECs, this study develops MAPCast, a GNN-based convection-allowing emulator, designed to generate surrogate background ensembles for high-resolution DA applications. MAPCast adopts an encoder-processor-decoder architecture operating on 7-level multiscale triangular meshes. The model is trained using historical MPAS simulations spanning 5 seasons during 2021-2025.

Surrogate background ensembles provided by MAPCast are evaluated using 10 retrospective convective storm cases in May 2018-2019 at forecast lead times of 15 and 60 min, corresponding to subhourly and hourly DA cycling. The primary findings are summarized as follows.

1. We first evaluate MAPCast in the short-term individual forecast members by comparing with MPAS forecasts. Results show that MAPCast reproduces short-term MPAS forecasts with good fidelity through the 60-min lead time, including realistic spatial and frequency distributions of state variables and similar spectral characteristics. Forecast skill decreases with increasing lead time for most variables. The primary discrepancies between MAPCast and MPAS are confined to small spatial scales, particularly near sharp gradients and convective-scale structures and for storm-related fields. 

2. We further evaluate MAPCast ensembles in the estimate of ensemble spread in a scale-dependent manner. Results show that MAPCast accurately reproduces both the magnitude and spatial distribution of large-scale MPAS ensemble spread across all evaluated variables. While MAPCast produces promising small-scale ensemble spread, substantially larger variability appears at the small scale, where spread statistics largely govern the behavior of the full-field results. As forecast lead time increases from 15 to 60 min, the agreement between MAPCast and MPAS spread decreases systematically. Nevertheless, large-scale spread structures remain substantially more robust than their full- and small-scale counterparts during forecast evolution. 

3. Final evaluation on MAPCast ensembles is performed to estimate univariate and multivariate spatial correlations. Results show that MAPCast reproduces MPAS univariate spatial correlations with a high fidelity and robustness for both large and small scale. The large-scale correlations keep a stronger agreement between MAPCast and MPAS ensembles across most variables, followed by the small-scale correlations. Similarly, the emulator most accurately captures the broad large-scale multivariate coupling structures relative to the corresponding full- and small-scale correlations. Despite that, MAPCast replicates the small-scale univariate and multivariate correlations better than their full-scale counterparts. Reproducing cross-variable covariance structures remains more challenging than reproducing single-variable correlations. Increasing forecast lead time from 15 to 60 min leads to systematic degradation in both univariate and multivariate correlation statistics.  

The baseline MAPCast configuration adopts architectural choices similar to those used in existing convection-allowing ML emulators, including a gridpoint-based MSE loss function, reduced vertical resolution, and an encoder-processor-decoder framework. The primary objective of the present evaluation is therefore not to optimize deterministic forecast skill, but to assess the capability of MAPCast to generate surrogate background ensembles suitable for estimating BECs. The results of this study indicate that further development of MAPCast is required to improve the representation of near-grid-scale structures and cross-variable covariance structures as well as the performance at longer lead times. Future work will target fine-scale error structures through scale-aware loss functions, like the spherical-harmonic decomposition of the MSE loss proposed by \cite{subichFixingDoublePenalty2025}. Improvements in multivariate covariance representation will also be explored using generative and diffusion-based approaches, such as those employed in GenCast \cite{priceProbabilisticWeatherForecasting2025}, which may better preserve physically consistent ensemble variability across variables.
The next phase of this work will incorporate MAPCast surrogate ensembles into a cycling DA framework to explore their direct impact on high-resolution DA. Kubalek et al. (2026) showed that using hybrid ensembles combining ML-based and physics-based members in ensemble DA outperform either pure ML or pure physics ensembles at equivalent computational cost within an idealized modeling framework. Motivated by this finding, future efforts will investigate combining MAPCast and MPAS background ensembles within the MPAS-JEDI DA system. Such integration requires MAPCast ensembles defined on the same vertical grid as MPAS, necessitating retraining it with full vertical resolution. In addition, the scale-dependent results presented in Sections 4.2 and 4.3 highlight the importance of scale-aware correlation treatment when incorporating surrogate ensembles into DA. In particular, the superior large- and small-scale covariance fidelity relative to full-scale performance suggests that surrogate ensembles may be especially well suited for simultaneous multiscale DA frameworks employing scale-dependent localization (SDL; \cite{wangSimultaneousMultiscaleData2023,wangImprovingCONUSConvectiveScale2024}). Future work will therefore focus on maximizing the advantage of surrogate background ensembles for multiscale BEC estimation within the convection-allowing MPAS-JEDI DA system.

\section*{Acknowledgements}
This study was supported by NOAA/Office of Oceanic and Atmospheric Research grant NA24OARX459C0001-T1-01. The authors would like to express their gratitude for the following computing resources. The historical MPAS simulations as the training dataset were generated using the computing resources provided by Orion, hosted in NOAA’s cooperative institute based at Mississippi State University. The MAPCast development and optimization, data storage, and surrogate ensemble evaluations were performed at the University of Oklahoma Supercomputing Center for Education and Research (OSCER). The MAPCast training was completed at the Texas Advanced Computing Center’s (TACC) Stampede3 System at the University of Texas at Austin, which is part of the Advanced Cyberinfrastructure Coordination Ecosystem (ACCESS). The authors acknowledge the discussions with Drs. Sergey Frolov, David Dowell, Montgomery Flora, and Corey Potvin. Additionally, we would also like to thank Dr. Yue Yang for her valuable comments, which significantly improved the initial version of this manuscript.

%\nolinenumbers

%This is where your bibliography is generated. Make sure that your .bib file is actually called library.bib
\bibliography{reference}

\begin{thebibliography}{10}

\bibitem{abdiHRRRCastDataDrivenEmulator2026}
D.~Abdi, I.~Jankov, P.~Madden, V.~Vargas, T.~A. Smith, S.~Frolov, M.~Flora, and C.~Potvin.
\newblock {{HRRRCast}}: {{A Data-Driven Emulator}} for {{Regional Weather Forecasting}} at {{Convection-Allowing Scales}}.
\newblock 5(2).

\bibitem{baLayerNormalization2016}
J.~L. Ba, J.~R. Kiros, and G.~E. Hinton.
\newblock Layer {{Normalization}}.

\bibitem{biAccurateMediumrangeGlobal2023}
K.~Bi, L.~Xie, H.~Zhang, X.~Chen, X.~Gu, and Q.~Tian.
\newblock Accurate medium-range global weather forecasting with {{3D}} neural networks.
\newblock 619(7970):533--538.

\bibitem{clarkRealTimeSimulatedForecasting2020}
A.~J. Clark, I.~L. Jirak, B.~T. Gallo, B.~Roberts, g.-i. family=Knopfmeier, given=Kent.~H., R.~A. Clark, J.~Vancil, A.~R. Dean, K.~A. Hoogewind, P.~L. Heinselman, N.~A. Dahl, M.~J. Krocak, J.~J. Choate, K.~A. Wilson, P.~S. Skinner, T.~A. Jones, Y.~Wang, G.~J. Creager, L.~J. Reames, L.~J. Wicker, S.~R. Dembek, and S.~J. Weiss.
\newblock A {{Real-Time}}, {{Simulated Forecasting Experiment}} for {{Advancing}} the {{Prediction}} of {{Hazardous Convective Weather}}.
\newblock 101(11):E2022--E2024.

\bibitem{dowellHighResolutionRapidRefresh2022}
D.~C. Dowell, C.~R. Alexander, E.~P. James, S.~S. Weygandt, S.~G. Benjamin, G.~S. Manikin, B.~T. Blake, J.~M. Brown, J.~B. Olson, M.~Hu, T.~G. Smirnova, T.~Ladwig, J.~S. Kenyon, R.~Ahmadov, D.~D. Turner, J.~D. Duda, and T.~I. Alcott.
\newblock The {{High-Resolution Rapid Refresh}} ({{HRRR}}): {{An Hourly Updating Convection-Allowing Forecast Model}}. {{Part I}}: {{Motivation}} and {{System Description}}.
\newblock 37(8):1371--1395.

\bibitem{dowellAdditiveNoiseStormScale2009}
D.~C. Dowell and L.~J. Wicker.
\newblock Additive {{Noise}} for {{Storm-Scale Ensemble Data Assimilation}}.
\newblock 26(5):911--927.

\bibitem{dudaComparingAssimilationRadar2019}
J.~D. Duda, X.~Wang, Y.~Wang, and J.~R. Carley.
\newblock Comparing the {{Assimilation}} of {{Radar Reflectivity Using}} the {{Direct GSI-Based Ensemble}}–{{Variational}} ({{EnVar}}) and {{Indirect Cloud Analysis Methods}} in {{Convection-Allowing Forecasts}} over the {{Continental United States}}.
\newblock 147(5):1655--1678.

\bibitem{fabryWhyAreRadar2020}
F.~Fabry and V.~Meunier.
\newblock Why {{Are Radar Data}} so {{Difficult}} to {{Assimilate Skillfully}}?
\newblock 148(7):2819--2836.

\bibitem{floraWoFSCastMachineLearning2025}
M.~L. Flora and C.~Potvin.
\newblock {{WoFSCast}}: {{A Machine Learning Model}} for {{Predicting Thunderstorms}} at {{Watch}}‐to‐{{Warning Scales}}.
\newblock 52(10):e2024GL112383.

\bibitem{flournoyInfluenceGroundRelativeFlow2021}
M.~D. Flournoy and E.~N. Rasmussen.
\newblock The {{Influence}} of {{Ground-Relative Flow}} and {{Friction}} on {{Near-Surface Storm-Relative Helicity}}.
\newblock 78(7):2135--2142.

\bibitem{gasperoniComparisonMethodsSample2020}
N.~A. Gasperoni, X.~Wang, and Y.~Wang.
\newblock A {{Comparison}} of {{Methods}} to {{Sample Model Errors}} for {{Convection-Allowing Ensemble Forecasts}} in the {{Setting}} of {{Multiscale Initial Conditions Produced}} by the {{GSI-Based EnVar Assimilation System}}.
\newblock 148(3):1177--1203.

\bibitem{guerretteDataAssimilationModel2023}
J.~J. Guerrette, Z.~Liu, C.~Snyder, B.-J. Jung, C.~S. Schwartz, J.~Ban, S.~Vahl, Y.~Wu, I.~H. Baños, Y.~G. Yu, S.~Ha, Y.~Trémolet, T.~Auligné, C.~Gas, B.~Ménétrier, A.~Shlyaeva, M.~Miesch, S.~Herbener, E.~Liu, D.~Holdaway, and B.~T. Johnson.
\newblock Data assimilation for the {{Model}} for {{Prediction Across Scales}} – {{Atmosphere}} with the {{Joint Effort}} for {{Data}} assimilation {{Integration}} ({{JEDI-MPAS}} 2.0.0-beta): Ensemble of {{3D}} ensemble-variational ({{En-3DEnVar}}) assimilations.
\newblock 16(23):7123--7142.

\bibitem{hakimDynamicalTestsDeep2024}
G.~J. Hakim and S.~Masanam.
\newblock Dynamical {{Tests}} of a {{Deep Learning Weather Prediction Model}}.
\newblock 3(3):e230090.

\bibitem{iaconoRadiativeForcingLonglived2008}
M.~J. Iacono, J.~S. Delamere, E.~J. Mlawer, M.~W. Shephard, S.~A. Clough, and W.~D. Collins.
\newblock Radiative forcing by long‐lived greenhouse gases: {{Calculations}} with the {{AER}} radiative transfer models.
\newblock 113(D13):2008JD009944.

\bibitem{jamesHighResolutionRapidRefresh2022}
E.~P. James, C.~R. Alexander, D.~C. Dowell, S.~S. Weygandt, S.~G. Benjamin, G.~S. Manikin, J.~M. Brown, J.~B. Olson, M.~Hu, T.~G. Smirnova, T.~Ladwig, J.~S. Kenyon, and D.~D. Turner.
\newblock The {{High-Resolution Rapid Refresh}} ({{HRRR}}): {{An Hourly Updating Convection-Allowing Forecast Model}}. {{Part II}}: {{Forecast Performance}}.
\newblock 37(8):1397--1417.

\bibitem{johnsonImpactsInitialCondition2024}
A.~Johnson and X.~Wang.
\newblock Impacts of {{Initial Condition Perturbation Blending}} in 10- and 40-{{Member Convection-Allowing Ensemble Forecasts}}.
\newblock 152(6):1421--1441.

\bibitem{johnsonComparisonMultiscaleGSIBased2015}
A.~Johnson, X.~Wang, J.~R. Carley, L.~J. Wicker, and C.~Karstens.
\newblock A {{Comparison}} of {{Multiscale GSI-Based EnKF}} and {{3DVar Data Assimilation Using Radar}} and {{Conventional Observations}} for {{Midlatitude Convective-Scale Precipitation Forecasts}}.
\newblock 143(8):3087--3108.

\bibitem{keislerForecastingGlobalWeather2022}
R.~Keisler.
\newblock Forecasting {{Global Weather}} with {{Graph Neural Networks}}.

\bibitem{lamLearningSkillfulMediumrange2023}
R.~Lam, A.~Sanchez-Gonzalez, M.~Willson, P.~Wirnsberger, M.~Fortunato, F.~Alet, S.~Ravuri, T.~Ewalds, Z.~Eaton-Rosen, W.~Hu, A.~Merose, S.~Hoyer, G.~Holland, O.~Vinyals, J.~Stott, A.~Pritzel, S.~Mohamed, and P.~Battaglia.
\newblock Learning skillful medium-range global weather forecasting.
\newblock 382(6677):1416--1421.

\bibitem{lawsonAdvancingConvectionAllowingNWP2018}
J.~R. Lawson, J.~S. Kain, N.~Yussouf, D.~C. Dowell, D.~M. Wheatley, K.~H. Knopfmeier, and T.~A. Jones.
\newblock Advancing from {{Convection-Allowing NWP}} to {{Warn-on-Forecast}}: {{Evidence}} of {{Progress}}.
\newblock 33(2):599--607.

\bibitem{liFuXiEn4DVarAssimilationSystem2024}
Y.~Li, W.~Han, H.~Li, W.~Duan, L.~Chen, X.~Zhong, J.~Wang, Y.~Liu, and X.~Sun.
\newblock {{FuXi}}‐{{En4DVar}}: {{An Assimilation System Based}} on {{Machine Learning Weather Forecasting Model Ensuring Physical Constraints}}.
\newblock 51(22):e2024GL111136.

\bibitem{lombardoProcessesControllingStructure2013}
K.~Lombardo and B.~A. Colle.
\newblock Processes {{Controlling}} the {{Structure}} and {{Longevity}} of {{Two Quasi-Linear Convective Systems Crossing}} the {{Southern New England Coast}}.
\newblock 141(11):3710--3734.

\bibitem{loshchilovDecoupledWeightDecay2019}
I.~Loshchilov and F.~Hutter.
\newblock Decoupled {{Weight Decay Regularization}}.

\bibitem{nakanishiDevelopmentImprovedTurbulence2009}
M.~Nakanishi and H.~Niino.
\newblock Development of an {{Improved Turbulence Closure Model}} for the {{Atmospheric Boundary Layer}}.
\newblock 87(5):895--912.

\bibitem{oskarssonGraphbasedNeuralWeather2023}
J.~Oskarsson, T.~Landelius, and F.~Lindsten.
\newblock Graph-based {{Neural Weather Prediction}} for {{Limited Area Modeling}}.

\bibitem{pathakKilometerscaleConvectionallowingModel2026}
J.~Pathak, Y.~Cohen, P.~Garg, P.~Harrington, N.~Brenowitz, D.~Durran, M.~Mardani, A.~Vahdat, S.~Xu, K.~Kashinath, and M.~Pritchard.
\newblock Kilometer-scale convection-allowing model emulation using generative diffusion modeling.
\newblock 12(5):eadv0423.

\bibitem{pathakFourCastNetGlobalDatadriven2022}
J.~Pathak, S.~Subramanian, P.~Harrington, S.~Raja, A.~Chattopadhyay, M.~Mardani, T.~Kurth, D.~Hall, Z.~Li, K.~Azizzadenesheli, P.~Hassanzadeh, K.~Kashinath, and A.~Anandkumar.
\newblock {{FourCastNet}}: {{A Global Data-driven High-resolution Weather Model}} using {{Adaptive Fourier Neural Operators}}.

\bibitem{priceProbabilisticWeatherForecasting2025}
I.~Price, A.~Sanchez-Gonzalez, F.~Alet, T.~R. Andersson, A.~El-Kadi, D.~Masters, T.~Ewalds, J.~Stott, S.~Mohamed, P.~Battaglia, R.~Lam, and M.~Willson.
\newblock Probabilistic weather forecasting with machine learning.
\newblock 637(8044):84--90.

\bibitem{ramachandranSearchingActivationFunctions2017}
P.~Ramachandran, B.~Zoph, and Q.~V. Le.
\newblock Searching for {{Activation Functions}}.

\bibitem{raspDataDrivenMediumRangeWeather2021}
S.~Rasp and N.~Thuerey.
\newblock Data‐{{Driven Medium}}‐{{Range Weather Prediction With}} a {{Resnet Pretrained}} on {{Climate Simulations}}: {{A New Model}} for {{WeatherBench}}.
\newblock 13(2):e2020MS002405.

\bibitem{robertsScaleSelectiveVerificationRainfall2008}
N.~M. Roberts and H.~W. Lean.
\newblock Scale-{{Selective Verification}} of {{Rainfall Accumulations}} from {{High-Resolution Forecasts}} of {{Convective Events}}.
\newblock 136(1):78--97.

\bibitem{scherDataDrivenWeatherClimate2018}
S.~Scher.
\newblock Toward {{Data}}‐{{Driven Weather}} and {{Climate Forecasting}}: {{Approximating}} a {{Simple General Circulation Model With Deep Learning}}.
\newblock 45(22).

\bibitem{schumacherConvectiveStormEnvironmentsSubtropical2021}
R.~S. Schumacher, D.~A. Hence, S.~W. Nesbitt, R.~J. Trapp, K.~A. Kosiba, J.~Wurman, P.~Salio, M.~Rugna, A.~C. Varble, and N.~R. Kelly.
\newblock Convective-{{Storm Environments}} in {{Subtropical South America}} from {{High-Frequency Soundings}} during {{RELAMPAGO-CACTI}}.
\newblock 149(5):1439--1458.

\bibitem{skamarockMultiscaleNonhydrostaticAtmospheric2012}
W.~C. Skamarock, J.~B. Klemp, M.~G. Duda, L.~D. Fowler, S.-H. Park, and T.~D. Ringler.
\newblock A {{Multiscale Nonhydrostatic Atmospheric Model Using Centroidal Voronoi Tesselations}} and {{C-Grid Staggering}}.
\newblock 140(9):3090--3105.

\bibitem{skinnerObjectBasedVerificationPrototype2018}
P.~S. Skinner, D.~M. Wheatley, K.~H. Knopfmeier, A.~E. Reinhart, J.~J. Choate, T.~A. Jones, G.~J. Creager, D.~C. Dowell, C.~R. Alexander, T.~T. Ladwig, L.~J. Wicker, P.~L. Heinselman, P.~Minnis, and R.~Palikonda.
\newblock Object-{{Based Verification}} of a {{Prototype Warn-on-Forecast System}}.
\newblock 33(5):1225--1250.

\bibitem{slivinskiAssimilatingObservedSurface2025}
L.~C. Slivinski, J.~S. Whitaker, S.~Frolov, T.~A. Smith, and N.~Agarwal.
\newblock Assimilating {{Observed Surface Pressure Into ML Weather Prediction Models}}.
\newblock 52(6):e2024GL114396.

\bibitem{smirnovaModificationsRapidUpdate2016}
T.~G. Smirnova, J.~M. Brown, S.~G. Benjamin, and J.~S. Kenyon.
\newblock Modifications to the {{Rapid Update Cycle Land Surface Model}} ({{RUC LSM}}) {{Available}} in the {{Weather Research}} and {{Forecasting}} ({{WRF}}) {{Model}}.
\newblock 144(5):1851--1865.

\bibitem{smithMultiRadarMultiSensorMRMS2016}
T.~M. Smith, V.~Lakshmanan, G.~J. Stumpf, K.~L. Ortega, K.~Hondl, K.~Cooper, K.~M. Calhoun, D.~M. Kingfield, K.~L. Manross, R.~Toomey, and J.~Brogden.
\newblock Multi-{{Radar Multi-Sensor}} ({{MRMS}}) {{Severe Weather}} and {{Aviation Products}}: {{Initial Operating Capabilities}}.
\newblock 97(9):1617--1630.

\bibitem{sodhiBenefitsSmoothingBackgrounds2022}
J.~S. Sodhi and F.~Fabry.
\newblock Benefits of {{Smoothing Backgrounds}} and {{Radar Reflectivity Observations}} for {{Multiscale Data Assimilation}} with an {{Ensemble Kalman Filter}} at {{Convective Scales}}: {{A Proof-of-Concept Study}}.
\newblock 150(3):589--601.

\bibitem{stensrudProgressChallengesWarnonForecast2013}
D.~J. Stensrud, L.~J. Wicker, M.~Xue, D.~T. Dawson, N.~Yussouf, D.~M. Wheatley, T.~E. Thompson, N.~A. Snook, T.~M. Smith, A.~D. Schenkman, C.~K. Potvin, E.~R. Mansell, T.~Lei, K.~M. Kuhlman, Y.~Jung, T.~A. Jones, J.~Gao, M.~C. Coniglio, H.~E. Brooks, and K.~A. Brewster.
\newblock Progress and challenges with {{Warn-on-Forecast}}.
\newblock 123:2--16.

\bibitem{stensrudConvectiveScaleWarnonForecastSystem2009}
D.~J. Stensrud, M.~Xue, L.~J. Wicker, K.~E. Kelleher, M.~P. Foster, J.~T. Schaefer, R.~S. Schneider, S.~G. Benjamin, S.~S. Weygandt, J.~T. Ferree, and J.~P. Tuell.
\newblock Convective-{{Scale Warn-on-Forecast System}}: {{A Vision}} for 2020.
\newblock 90(10):1487--1500.

\bibitem{subichFixingDoublePenalty2025}
C.~Subich, S.~Z. Husain, L.~Separovic, and J.~Yang.
\newblock Fixing the {{Double Penalty}} in {{Data-Driven Weather Forecasting Through}} a {{Modified Spherical Harmonic Loss Function}}.

\bibitem{thompsonStudyAerosolImpacts2014}
G.~Thompson and T.~Eidhammer.
\newblock A {{Study}} of {{Aerosol Impacts}} on {{Clouds}} and {{Precipitation Development}} in a {{Large Winter Cyclone}}.
\newblock 71(10):3636--3658.

\bibitem{wangCosteffectiveHybridEnVar2026}
Y.~Wang and X.~Wang.
\newblock Cost‐effective hybrid {{{\textsc{EnVar}}}} for convective‐scale data assimilation through machine‐learning‐aided background error covariances.
\newblock 152(776):e70152.

\bibitem{wangDirectAssimilationRadar2017}
Y.~Wang and X.~Wang.
\newblock Direct {{Assimilation}} of {{Radar Reflectivity}} without {{Tangent Linear}} and {{Adjoint}} of the {{Nonlinear Observation Operator}} in the {{GSI-Based EnVar System}}: {{Methodology}} and {{Experiment}} with the 8 {{May}} 2003 {{Oklahoma City Tornadic Supercell}}.
\newblock 145(4):1447--1471.

\bibitem{wangImprovingCONUSConvectiveScale2024}
Y.~Wang and X.~Wang.
\newblock Improving {{CONUS Convective}}‐{{Scale Forecasting With Simultaneous Multiscale Data Assimilation}}.
\newblock 129(14):e2023JD040206.

\bibitem{wangMultivariateAdditiveInflation2023}
Y.~Wang and X.~Wang.
\newblock A {{Multivariate Additive Inflation Approach}} to {{Improve Storm}}‐{{Scale Ensemble}}‐{{Based Data Assimilation}} and {{Forecasts}}: {{Methodology}} and {{Experiment With}} a {{Tornadic Supercell}}.
\newblock 15(1):e2022MS003307.

\bibitem{wangSimultaneousMultiscaleData2023}
Y.~Wang and X.~Wang.
\newblock Simultaneous {{Multiscale Data Assimilation Using Scale}}‐ and {{Variable}}‐{{Dependent Localization}} in {{EnVar}} for {{Convection Allowing Analyses}} and {{Forecasts}}: {{Methodology}} and {{Experiments}} for a {{Tornadic Supercell}}.
\newblock 15(5):e2022MS003430.

\bibitem{weynImprovingDataDrivenGlobal2020}
J.~A. Weyn, D.~R. Durran, and R.~Caruana.
\newblock Improving {{Data}}‐{{Driven Global Weather Prediction Using Deep Convolutional Neural Networks}} on a {{Cubed Sphere}}.
\newblock 12(9):e2020MS002109.

\bibitem{whitakerEvaluatingMethodsAccount2012}
J.~S. Whitaker and T.~M. Hamill.
\newblock Evaluating {{Methods}} to {{Account}} for {{System Errors}} in {{Ensemble Data Assimilation}}.
\newblock 140(9):3078--3089.

\bibitem{workoffInfluenceLakeErie2012}
T.~E. Workoff, D.~A.~R. Kristovich, N.~F. Laird, R.~LaPlante, and D.~Leins.
\newblock Influence of the {{Lake Erie Overlake Boundary Layer}} on {{Deep Convective Storm Evolution}}.
\newblock 27(5):1279--1289.

\bibitem{yangImpactRadarReflectivity2023}
Y.~Yang and X.~Wang.
\newblock Impact of {{Radar Reflectivity Data Assimilation Frequency}} on {{Convection-Allowing Forecasts}} of {{Diverse Cases}} over the {{Continental United States}}.
\newblock 151(2):341--362.

\bibitem{zhangImpactsInitialEstimate2004}
F.~Zhang, C.~Snyder, and J.~Sun.
\newblock Impacts of {{Initial Estimate}} and {{Observation Availability}} on {{Convective-Scale Data Assimilation}} with an {{Ensemble Kalman Filter}}.
\newblock 132(5):1238--1253.

\end{thebibliography}

%This defines the bibliographies style. Search online for a list of available styles.
\bibliographystyle{abbrv}

\end{document}